\documentclass[onecolumn,draftclsnofoot]{IEEEtran}

\usepackage{latexsym}
\usepackage{amssymb}
\usepackage{bm}
\usepackage{stmaryrd}
\usepackage[dvips]{graphics}
\usepackage{graphicx}
\usepackage{psfrag}
\usepackage{amsfonts,amsmath,amssymb,amsthm}
\usepackage{algorithm}
\usepackage{algorithmic}
\usepackage{dsfont,color,subfigure,setspace}
\usepackage{cite,pdfsync}
\usepackage{float,cite}
\DeclareMathOperator*{\argmin}{arg\,min}

\usepackage{xspace}
\usepackage{bbm}

\newcommand{\Ac}{\mathcal{A}}
\newcommand{\Bc}{\mathcal{B}}
\newcommand{\Cc}{\mathcal{C}}

\newcommand{\Ec}{\mathcal{E}}
\newcommand{\Fc}{\mathcal{F}}

\newcommand{\Nc}{\mathcal{N}}

\newcommand{\Pc}{\mathcal{P}}

\newcommand{\Sc}{\mathcal{S}}
\newcommand{\Tc}{\mathcal{T}}
\newcommand{\Uc}{\mathcal{U}}
\newcommand{\Vc}{\mathcal{V}}

\newcommand{\Xc}{\mathcal{X}}
\newcommand{\Yc}{\mathcal{Y}}
\newcommand{\Zc}{\mathcal{Z}}

\newcommand{\Xh}{{\hat{X}}}

\newcommand{\xh}{{\hat{x}}}

\newcommand{\Et}{{\tilde{E}}}

\newcommand{\xt}{{\tilde{x}}}

\def\a{\alpha}

\def\g{\gamma}
\def\d{\delta}
\def\e{\epsilon}

\def\l{\lambda}

\DeclareMathOperator\Ex{E}
\let\P\relax
\DeclareMathOperator\P{P}

\newcommand{\Bern}{\mathrm{Bern}}

\newcommand\ie{i.e.,\xspace}
\def\textiid{i.i.d.\@\xspace}
\newcommand\iid{\ifmmode\text{ i.i.d. } \else \textiid \fi}

\newcommand{\ind}{\mathbbmss{1}}

\newcommand{\ex}{{\rm e}}

\newcommand{\Xbbf}{{\mathbb{\mathbf{X}}}}
\newcommand{\Ubbf}{{\mathbb{\mathbf{U}}}}
\newcommand{\Zbbf}{{\mathbb{\mathbf{Z}}}}
\newtheorem{definition}{Definition}
\newtheorem{theorem}{Theorem}
\newtheorem{lemma}{Lemma}

\newtheorem{corollary}{Corollary}
\newtheorem{remark}{Remark}

\begin{document}

\title{New approach to Bayesian high-dimensional linear regression}
\author{
    \IEEEauthorblockN{Shirin Jalali, Arian Maleki}}

\maketitle

\begin{abstract}
Consider the problem of estimating parameters $X^n \in \mathbb{R}^n $, generated by a stationary process, from $m$ response variables $Y^m = AX^n+Z^m$, under the assumption that the distribution of $X^n$ is known. This is the most general version of the Bayesian linear regression  problem. The lack of computationally feasible algorithms that can employ generic prior distributions and provide a good estimate of $X^n$ has limited the set of distributions researchers use to model the data. In this paper, a new  scheme called Q-MAP is proposed. The new method has the following properties: (i) It has similarities to the popular MAP estimation under the noiseless setting. (ii) In the noiseless setting, it achieves the ``asymptotically optimal performance'' when $X^n$ has independent and identically distributed components. (iii) It scales favorably with the dimensions of the problem and therefore is applicable to high-dimensional setups. (iv) The solution of the Q-MAP optimization can be found via a proposed iterative algorithm which is provably robust to the error (noise) in the response variables. 
\end{abstract}


\section{Introduction}

\subsection{Motivation}

Consider the problem of linear regression under the Bayesian setup; The parameter vector of length $n$ denoted by $X^n$ is generated by a stationary stochastic  process  $\Xbbf=\{X_i\}_{1}^{\infty}$, whose distribution is known. The goal is to estimate $X^n$ from  $m<n$  response variables of the form $Y^m=AX^n + Z^m$, where $A\in\mathds{R}^{m\times n}$ and  $Z^m\in\mathds{R}^m$ denote the design matrix and the noise vector, respectively.  (The case of $m\geq n$ is also of interest and is valid under our model.) In order to solve this problem, there are two  fundamental questions that can be raised and are studied in this paper:
\begin{enumerate}
\item How should we use the distribution of $X^n$ to obtain an efficient estimator? In order to answer this question, there are two main criteria that should be taken into account: (i) computational complexity: how efficiently can the estimate be computed? (ii) accuracy: how well does the estimator perform? If we ignore the computational complexity constraint, then the answer to our first question is simple. An optimal Bayes estimator seeks to minimize the Bayes risk defined as $\Ex[\ell(\hat{X}^n, X^n)]$, where $\ell: \mathds{R}^n\times\mathds{R}^n\to\mathds{R}^+$ denotes the considered  loss function. For instance,  $\ell(x^n,\xh^n)=\|x^n-\xh^n\|_2^2$ leads to the  minimum mean square error (MMSE) estimator. However, the computational complexity of calculating the MMSE estimation for generic distributions is very high. 

\item Can the performance of the estimator be analyzed in high-dimensional settings?  The answer to this question is also complicated. Even the performance analysis of standard estimators such as the MMSE estimator is challenging. In fact, even if we assume that $\mathbf{X}$ is an independent and identically distributed process, the analysis is still complicated and heuristic tools such as replica method from statistical physics have been employed to achieve this goal.  
\end{enumerate}

In response to the above two questions we propose an optimization problem, which we refer to as  quantized maximum a posteriori (Q-MAP). We then show how this optimization problem can be analyzed and solved. Before presenting the Q-MAP optimization, we review some required notations.  

\subsection{Notation}

Calligraphic letters such as $\Xc$ and $\Yc$ denote sets. The size of a set $\Xc$ is denoted by $|\Xc|$. Given vector $(v_1,v_2,\ldots,v_n)\in\mathds{R}^n$ and integers $i,j\in\{1,\ldots,n\}$, where $i\leq j$, $v_i^j\triangleq (v_i,v_{i+1},\ldots,v_j).$ 
For simplicity $v_1^j$ and $v_j^j$ are denoted by $v^j$ and $v_j$, respectively. For two vectors $u^n$ and $v^n$, both in $\mathds{R}^n$, let $\langle u^n,v^n\rangle$ denote their inner product defined as $\langle u^n,v^n\rangle\triangleq \sum_{i=1}^nu_iv_i$. The all-zero and all-one vectors of length $n$ are denoted by $0^n$ and $1^n$, respectively.

Uppercase letters such as $X$ and $Y$ denote random variables. The alphabet of a random variable $X$ is denoted by $\Xc$. The entropy \cite{cover} of a finite-alphabet random variable $U$ with probability mass function (pmf) $p(u)$, $u\in\Uc$, is defined as
$H(U)=-\sum_{u\in\Uc}p(u)\log{ p(u)}.$ 
Given finite-alphabet random variables $U$ and $V$ with joint pmf $p(u,v)$, $(u,v)\in\Uc\times \Vc$, the conditional entropy of $U$ given $V$ is defined as $H(U|V)=-\sum_{(u,v)\in\Uc\times \Vc} p(u,v)\log p(u|v).$

Matrices are also denoted by uppercase letters such as $A$ and $B$ and are differentiated from random variables by context.  Throughout the paper $\log$ and $\ln$ refer to logarithm in base 2 and the natural logarithm, respectively.

For $x\in\mathds{R}$, $\lfloor x\rfloor$ denotes the largest integer smaller than $x$. Therefore, $0\leq x-\lfloor x\rfloor<1$, for all $x$. The $b$-bit quantized version of $x$ is denoted by $[x]_b$ and is defined as 
\begin{align}
[x]_b\triangleq\lfloor x\rfloor+\sum_{i=1}^b2^{-i}a_i,\end{align} 
where for all $i$, $a_i\in\{0,1\}$, and $0.a_1a_2\ldots$ denotes the binary representation of $x-\lfloor x\rfloor$.  For a vector $x^n\in\mathds{R}^n$, $[x^n]_b\triangleq([x_1]_b,\ldots,[x_n]_b).$

Consider a vector $u^n\in\Uc^n$, where $|\Uc|<\infty$.  The $(k+1)^{th}$ order empirical distribution of $u^n$ is denoted by $\hat{p}^{(k+1)}$, and is defined as follows: for any $a^{k+1}\in \Uc^{k+1}$,
\begin{align}
\hat{p}^{(k+1)}(a^{k+1}| u^n)&\triangleq {|\{i: u_{i-k}^i=a^{k+1}, k+1\leq i\leq n\}|\over n-k}\nonumber\\
&={1\over n-k}\sum_{i=k+1}^n\ind_{u_{i-k}^{i}=a^{k+1}},\label{eq:emp-dist}
\end{align}
where $\ind_{\Ec}$ denotes the indicator function of event $\Ec$. 

\subsection{Contributions}\label{ssec:contributions}

Consider the stochastic process $\Xbbf=\{X_i\}_{i=1}^{\infty}$ discussed earlier. Let $X_i\in\Xc$, for all $i$.  Assume that  $\Xc$ is a bounded subset of $\mathds{R}$. Define the $b$-bit quantized version of $\Xc$ as
\begin{align}
\Xc_b\triangleq \{[x]_b: \;x\in\Xc\}.
\end{align}  
Note that if $\Xc$ is a bounded set, then $\Xc_b$ is a finite set, \ie $|\Xc_b|<\infty$.
The Q-MAP optimization estimates $X^n$ from $Y^m=AX^n+Z^m$, by solving the following optimization:
\begin{align}\label{eq:Q-MAP}
  \Xh^n&\;\;= \;\; \argmin_{u^n\in\Xc_b^n} \;\;\;\|Au^n-Y^m\|^2 \nonumber \\
& \;\;\;\;\;\; \;\;\;{\rm subject \ to}\  \sum_{a^{k+1}\in\Xc_b^{k+1}} w_{a^{k+1}} \hat{p}^{(k+1)}(a^{k+1}|u^n) \leq \gamma_n,
  \end{align}
where $\gamma_n$ is a number that may depend on $n$ and the distribution of $\Xbbf$, and  the non-negative weights $(w_{a^{k+1}}:\;a^{k+1}\in \Xc_b^{k+1})$ are  defined as a function of the probability distribution of the stochastic process $\Xbbf$ as:
\begin{align}
w_{a^{k+1}}\triangleq \log {1\over \P([X_{k+1}]_b=a_{k+1}|[X^k]_b=a^k)}.\label{eq:coeffs-Q-MAP}
\end{align}
 Note that $b$, $k$ and $\gamma_n$ are three parameters that have to be set properly. Proper choices for these three parameters will be described in our analysis.  

Note that minimizing $\|Au^n - Y^m\|^2$ is natural since we would like to obtain a parameter vector that matches the response variables. However, with no constraint on potential solution vectors the estimate will suffer from overfitting (unless $m$ is much larger than $n$). Hence, some constraints should be imposed on  the set of potential solutions. In Q-MAP optimization, this constraint requires a potential solution $u^n\in\Xc_n^n$ to satisfy 
\begin{align}
\sum_{a^{k+1}\in\Xc_b^{k+1}} w_{a^{k+1}} \hat{p}^{(k+1)}(a^{k+1}|u^n) \leq \gamma_n.
\end{align}
 As it will be described later,  the function 
 \begin{align}
 c_w(u^n)\triangleq \sum_{a^{k+1}\in\Xc_b^{k+1}} w_{a^{k+1}} \hat{p}^{(k+1)}(a^{k+1}|u^n)
 \end{align} is a measure  of ``complexity'' of the sequences in $\Xc_b^n$. For instance, as will be shown later in Section \ref{sec:iid-sparse},  for an independent identically distributed (i.i.d.) process $\Xbbf$ with $X_i\sim (1- p)\delta_0+ pf_c$, where $p\in(0,1)$, $\delta_0$ denotes a point mass at zero, and $f_c$ denotes an absolutely continuous distribution over a bounded set, the bound  $c_w(u^n)\leq \gamma_n$ with $k=0$ simplifies to a bound on the $\ell_0$-norm of sequence $u^n$.  Hence, intuitively-speaking, the constraint $ c_w(u^n) \leq \gamma_n$ ensures that the Q-MAP optimization only considers ``low-complexity'' sequences that comply with the known source model. There are two other features of the above optimization that are worth more emphasis and clarification at this point:
\begin{enumerate}
\item Quantized reconstructions: While the parameter vector $X^n$ and the response variables $Y^m$ are typically real valued, the estimate produced by the Q-MAP optimization lies in the quantized space $\Xc_b^n$. The motivation for this quantization will be explained  later, but in a nutshell, this step helps both the theoretical analysis and also the implementation of the optimization. Note that given the fact that the distributions are generic distributions, even storing the source model in a computer requires some type of quantization. Also, in certain applications we would like to learn the source distribution from a data-base. Again in those cases, quantization may be a natural step as it is done in the evaluation of histograms.

\item Memory parameter ($k$): again  both for the convenience of the theoretical analysis and also for the ease of implementation, only dependencies captured by the $(k+1)$-th order probability distribution of the process $\Xbbf$ are taken into account in the Q-MAP optimization. This memory parameter is an arbitrary free parameter that can be selected based on the source distribution. As shown later, for instance, in the noiseless setting, for an i.i.d.~process, $k=0$ is  enough to achieve the fundamental limits in terms of the minimum number of response variables.
\end{enumerate}


While the Q-MAP optimization provides a new approach to Bayesian compressed sensing, it is still not   an easy optimization problem. For instance, for the i.i.d.~distribution mentioned earlier $(1- p)\delta_0+ p f_c$,  the constraint becomes equivalent to having an upper bound on $\|u^n\|_0$. This is  similar to the notoriously difficult optimal variable selection problem. Hence, despite the fact that the Q-MAP optimization problems appear simpler than other estimators such as MMSE, it in fact can still be computationally infeasible. However, inspired by the projected gradient descent (PGD) method in convex optimization, we propose the following algorithm to solve the Q-MAP optimization. Define 
\begin{align}
\Fc_o\triangleq \Big\{u^n\in\Xc_b^n: \;  \sum_{a^{k+1}\in\Xc_b^{k+1}} w_{a^{k+1}}\hat{p}^{(k+1)}(a^{k+1}|u^n) \leq \gamma_n \Big\}.\label{eq:def-Fc-rand}
\end{align}
Note that the set $\Fc_o$ depends on quantization level $b$, memory parameter $k$, weights $\{w_{a^{k+1}}\}_{a^{k+1}\in\Xc_b^{k+1}}$ and also parameter  $\gamma_n$. For simplicity, these dependencies are not  explicitly shown in the expression of $\Fc_o$ as $\Fc_o(b,k,$ $\{w_{a^{k+1}}\}_{a^{k+1}\in\Xc_b^{k+1}},\gamma_n)$. 

The PGD algorithm generates a sequence of estimates $\Xh^n(t)$, $t=0,1,\ldots$ of the sequence $X^n$. It starts by setting $\Xh^n(0)=0^n$, and proceeds by updating $\Xh^n(t)$, its estimate at time $t$, as follows
\begin{align}
S^n(t+1)&=\Xh^n(t)+\mu A^T(Y^m-A\Xh^n(t))\nonumber\\
\Xh^n(t+1)&=\argmin_{u^n\in\Fc_o}\left\|u^n-S^n(t+1)\right\|,\label{eq:PGD-update}
\end{align}
where $\mu$ denotes the step-size and ensures that the algorithm does {\em not} diverge to infinity. Intuitively, the above procedure, at each step, moves the current estimate towards the $Ax^n=Y^m$ hyperplane and then projects the new estimate to the set of low-complexity vectors.   As will be discussed later, when $m$ is large enough, in the noiseless setting, the estimates provided by the PGD algorithm converge to $X^n$, with high probability.

The challenging  step in running the PGD method  is  the projection step, which requires  finding the closest point  to $S^n(t+1)$ in $\mathcal{F}_o$. For some special distributions such as sparse or piecewise-constant  discussed in Section  \ref{sec:special-dist}, the corresponding set $\Fc_o$ has a special form  that simplifies the projection considerably. In general, while  projection to a non-convex discrete set can be  complicated,  as we will discuss in Section \ref{sec:comp-comp}, we believe that because of the special  structure of the set $\mathcal{F}_o$, a dynamic programming approach can be used for performing this projection. More specifically,  we will explain how a  Viterbi algorithm \cite{viterbi1967error} with $2^{bk}$ states and $n$ stages can be used for this purpose. Hence,  the complexity of the proposed method for doing the projection task required by the PGD grows linearly in  $n$, but exponentially in $kb$. We expect that for ``structured distributions" the scaling with $b$ and $k$ can be improved much beyond this. We will describe our intuition in Section \ref{sec:comp-comp}, but leave the formal exploration  of this direction to future research.

The main question we have not addressed yet is  how well the Q-MAP optimization  and the proposed PGD method perform.  In the next few paragraphs, we informally state our main results. The upper information dimension of a stationery process,  defined as \cite{JalaliP:14-arxiv}, plays a key role in our analysis. The $k$-th order upper information dimension of stationary process $\Xbbf$ is defined as 
\begin{equation}\label{eq:first_appearanced_k}
\bar{d}_k(\Xbbf)\triangleq \limsup_{b\to \infty} {H([X_{k+1}]_b|[X^k]_b) \over b},
\end{equation}
where $H([X_{k+1}]_b|[X^k]_b) $ denotes the  conditional entropy of $[X_{k+1}]_b$ given $[X^k]_b$. Similarly, the lower $k$-th order upper information dimension of  $\Xbbf$ is defined as 
\begin{align}
\underline{d}_k(\Xbbf)\triangleq \liminf_{b\to \infty} {H([X_{k+1}]_b|[X^k]_b) \over b}.
\end{align}
If $\bar{d}_k(\Xbbf)=\underline{d}_k(\Xbbf)$, then the   $k$-th order  information dimension of process  $\Xbbf$ is defined as ${d}_k(\Xbbf)=\bar{d}_k(\Xbbf)=\underline{d}_k(\Xbbf)$.
  For $k=0$, $\bar{d}_k(\Xbbf)$ ($\underline{d}_k(\Xbbf)$) is equal to the upper (lower) R\'enyi information  dimension of $X_1$ \cite{Renyi:59}, which is a well-known measure of complexity  for random variables or random vectors. Also, it can be proved that for all stationary sources with $H(\lfloor X_1 \rfloor)$, $\bar{d}_k(\Xbbf)\leq 1$ and $\underline{d}_k(\Xbbf)\leq 1$ \cite{JalaliP:14-arxiv}.
  
  To gain some insight on these defenitions, consider an i.i.d. process $\Xbbf$ with $X_1 \sim (1-p) \delta_0(x) + p {\rm Unif}(0,1)$, where  ${\rm Unif}$ denotes a uniform distribution. This is called the spike and slab prior \cite{mitchell1988bayesian}. It can be proved that for this process ${d}_0(\Xbbf)=\bar{d}_0(\Xbbf)=\underline{d}_0(\Xbbf) =p$ \cite{Renyi:59}. For general stationary sources with infinite memory, the limit of $\bar{d}_k(\Xbbf)$ as $k$ grows to infinity is defined as the upper information dimension of the process $\Xbbf$ and is denoted by $\bar{d}_o(\Xbbf)$. As argued in \cite{JalaliP:14-arxiv}, the information dimension of a process measures the ``complexity'' or the ``structure'' present in a process. Based on these definitions and concepts, we state our results in the following. The exposition of our results is informal and lacks many details. All the details will be clarified later in the paper.
  \vspace{.2cm}

\textbf{Informal result 1. } Consider the noiseless setting ($Z^m=0^m$), and assume that the elements of the desgin matrix $A$ are i.i.d.~Gaussian. Further assume that the process $\Xbbf$ satisfies certain mixing conditions, and for a fixed $k$, ${m\over n}> \bar{d}_k(\Xbbf)$. Then, asymptotically, for a proper quantization level which grows with $n$, the Q-MAP optimization  recovers $X^n$ with high probability. 

\vspace{.2cm}

There is an interesting feature of this result that we would like to emphasize here: (i) If $ \bar{d}_k(\Xbbf)$ is strictly smaller than $1$, then we can estimate $X^n$ accurately, from $m <n$ response variables. In fact the smaller  $\bar{d}_k(\Xbbf)$, the less response variables  are required. In particular, we can consider the spike and slab prior we discussed before that corresponds to sparse parameter vectors that are studied in the literature \cite{bickel2009simultaneous, candes2007dantzig}. For this prior $\bar{d}_0(\Xbbf) =p$. Hence, as long as $m> np$, asymptotically,  the estimate of Q-MAP with $k=0$ will be accurate. Note that $np$ is in fact the expected number of non-zero elements in $\beta$.

We believe that even an MMSE estimator that employs only the $k^{th}$ order distribution of the source cannot recover with a smaller number of response variables. We present some examples that confirm our claim, however the optimality of the result we obtain above is an open question that we leave for future research. Note that the above result is for Q-MAP that is still computationally complicated. Our next result is about our PGD-based algorithm. 
\vspace{.2cm}

\textbf{Informal result 2.} Consider again the noiseless setting, and assume that the elements of $A$ are i.i.d.~Gaussian. If the process $\Xbbf$ satisfies certain mixing conditions, and  ${m\over n}> 10 b \bar{d}_k(\Xbbf)$, where $k$ is a fixed parameter, then the estimates derived by the PGD algorithm, with high probability, converge to $X^n$.   We will also characterize the convergence rate of the PGD-based algorithm and  its performance in the presence of an additive white Gaussian noise (AWGN). 

Compared to Informal Result 1, the number of response variables required in Informal Result 2 is a factor $10b$ higher. As we will discuss later we let $b$ grow as in $O(\log \log n)$, and hence the difference between Informal Result 1 and Informal Result 2 is not substantial. 

 \subsection{Related work and discussion}
 
 Bayesian linear regression has been a topic of extensive research in the past fifty years \cite{mitchell1988bayesian, west1984outlier, park2008bayesian, HeCa09, som2012compressive, hoadley1970bayesian, liu1996bayesian, lindley1972bayes, tipping2001sparse, hans2009bayesian, hans2010model, o2009review}. In all these papers $X^n$ is considered as a random vector whose distribution is known. However, often simple models are considered for the distribution of $X^n$ to simplify either  the posterior calculations or to apply Markov chain Monte Carlo methods such as Gibbs sampling. This paper considers a different scenario. We ignore the computational issues at first and consider an arbitrary distribution for $X^n$. This is in particular useful for applications in which complicated prior can be learned. (For instance, one might have access to a  large database that has many different draws of process $\mathbf{X}$.) We then present an optimization for estimating $X^n$ and prove the optimality of this approach under some conditions. This approach let us avoid the limitations that are imposed by posterior calculations. On the other hand, one main advantage of having posterior distributions is that they can be used in calculating confidence intervals. Exploring confidence intervals and related topics  remains an open question for future research. 
 
Our theoretical analyses are inspired by the recent surge of interest toward understanding the high-dimensional linear regression problem \cite{CaTa05, candes2007dantzig, bickel2009simultaneous, rigollet2011exponential, su2015slope, donoho2013high}. In this front, there has been very limited work on the theoretical analysis of the Bayesian recovery algorithms, especially beyond memoryless sources. Two of the main tools that have been used for this purpose are the replica method \cite{guo2005randomly} and state evolution \cite{MalekiThesis}. Both methods  have been employed to analyze the asymptotic performance of MMSE and MAP estimators under the asymptotic setting where $m,n \rightarrow \infty$, while $m/n$ is fixed. They both have some limitations. For instance, replica-based methods are  not fully rigorous. Moreover, while they work well  for i.i.d.~sequences, it is not clear how they can be applied to sources with memory. The state evolution framework suffers from  similar issues. Our paper presents the first result in this direction for processes with memory.

%
%

\subsection{Organization of the paper}
The organization of the paper is as follows. The Q-MAP estimator   is developed in Section \ref{sec:Q-MAP}, and in Section \ref{sec:special-dist} it   is simplified for some simple distributions and shown to have connections to some well-known algorithms. In Section \ref{sec:exp-rate}, two classes of stochastic  processes are studied. The empirical distributions  of the quantized versions of processes in each class have  exponential convergence rates.   The performance of  the Q-MAP estimator is studied in Section \ref{sec:analysisqmap}. An iterative method based on PGD is  proposed and studied  in \ref{sec:solving-QMAP}.   Section \ref{sec:proof}   presents the proofs of the main results of the paper,  and finally Section  \ref{sec:conclusions} concludes the paper.


\section{Quantized MAP  estimator} \label{sec:Q-MAP}

Consider $X^n$ generated by a stationary process $\Xbbf$. Assume that there is no noise in the response variables, i.e., $Y^m=AX^n$. The goal is to estimate $X^n$ from $Y^m$ and $A$. Since there is no noise, we can employ a MAP estimator that  finds the most likely parameter vector for response variables $Y^m$. Instead of solving the original MAP estimator, we consider finding the most probable sequence in the quantized space $\Xc_b^n$. That is,
\begin{align}
{\rm maximize} \;\;&\P([X^n]_b=u^n)\nonumber\\
{\rm subject\; to}\;\; & u^n\in\Xc_b^n,\nonumber\\
&[x^n]_b=u^n,\nonumber\\
&Ax^n=Y^m,\label{eq:Q-MAP-orig}
\end{align} 
where $\P$ denotes the law of process $\mathbf{X}$. Note that when the source distribution does not have a specific parametric form, for implementation purposes, one should use quantization. Hence, quantization can be considered as a natural step in    implementation.

We  further simplify the  formulation in \eqref{eq:Q-MAP-orig} to make it more amenable to both analysis and implementation. Note that if $Ax^n=Y^m$, and $|x_i-u_i|\leq 2^{-b}$, for all $i$, then
\begin{align}
\|Au^n-Y^m\|_2^2&\leq (\sigma_{\max}(A))^2 \|x^n-u^n\|^2\nonumber\\
&\leq 2^{-2b}n^2(\sigma_{\max}(A))^2 ,\label{eq:Q-MAP-orig-simlifies-s1}
\end{align}
where $\sigma_{\rm max}(A)$ denotes the maximum singular value of the design matrix $A$.
This provides an upper bound on ${1\over n^2}\|A[x^n]_b-Y^m\|_2^2$ in terms of $\sigma_{\max}(A)$ and $b$.  In other words, since $u^n$ is a quantized version of $x^n$, and $Ax^n=Y^m$, $\|Au^n-Y^m\|_2^2$ is also expected to be small. 

In order to further simplify \eqref{eq:Q-MAP-orig-simlifies-s1}, we focus on the other term in \eqref{eq:Q-MAP-orig-simlifies-s1}, \ie $-\log\P([X^n]_b=u^n)$. Assume that the process $\Xbbf$ is such that $\P([X^n]_b=u^n)$ can be factored as 
\begin{align}
\P([X^n]_b=u^n)= \P([X^k]_b=u^k ) \prod_{i=k+1}^n\P([X_i]_b=u_i|[X_{i-k}^{i-1}]_b=u_{i-k}^{i-1}),
\end{align}
for some finite $k$. In other words, the $b$-bit quantized version of $\Xbbf$ is a Markov process of order $k$. Then define coefficients $(w_{a^{k+1}}:\;a^{k+1}\in \Xc_b^{k+1})$ according to \eqref{eq:coeffs-Q-MAP}. This assumption simplifies the term $-\log\P([X^n]_b=u^n)$ in the  following way:
\begin{align}
&-\log \P([X^n]_b=u^n)\nonumber\\
&=-\log \P([X^k]_b=u^k ) -\sum_{i=k+1}^n\log \P([X_i]_b=u_i|[X_{i-k}^{i-1}]_b=u_{i-k}^{i-1})\nonumber\\
&=-\log \P([X^k]_b=u^k ) \nonumber\\
&\;\;\;\;\;-\sum_{i=k+1}^n\log \P([X_i]_b=u_i|[X_{i-k}^{i-1}]_b=u_{i-k}^{i-1})\sum_{a^{k+1}\in \Xc_b^{k+1}}\ind_{u_{i-k}^i=a^{k+1}}\nonumber\\
&=-\log \P([X^k]_b=u^k ) \nonumber\\
&\;\;\;\;\;-\sum_{i=k+1}^n \sum_{a^{k+1}\in \Xc_b^{k+1}}\ind_{u_{i-k}^i=a^{k+1}}\log\P([X_i]_b=a_i|[X_{i-k}^{i-1}]_b=a_{i-k}^{i-1})\nonumber\\
&=-\log \P([X^k]_b=u^k ) + \sum_{i=k+1}^n \sum_{a^{k+1}\in \Xc_b^{k+1}}w_{a^{k+1}}\ind_{u_{i-k}^i=a^{k+1}}\nonumber\\
&=-\log \P([X^k]_b=u^k ) + \sum_{a^{k+1}\in \Xc_b^{k+1}}w_{a^{k+1}}\sum_{i=k+1}^n\ind_{u_{i-k}^i=a^{k+1}}\nonumber\\
&=-\log \P([X^k]_b=u^k ) + (n-k)\sum_{a^{k+1}\in \Xc_b^{k+1}} w_{a^{k+1}}\hat{p}^{(k+1)}(a^{k+1}|u^n),
\end{align}
where $\hat{p}^{(k+1)}$ is  the $(k+1)^{th}$ order empirical distribution of $u^n$ as defined in \eqref{eq:emp-dist}. Assuming $k$ is much smaller than $n$, and ignoring the negligible term of $-\log \P([X^k]_b=u^k )/(n-k)$,  instead of minimizing $-\log\P([X^n]_b=u^n)$, subject to an upper bound on  $\|Au^n-Y^m\|_2^2$, we consider the following optimization where the roles of the cost and constraint functions are flipped 
\begin{align}\label{eq:Q-MAP}
  \Xh^n&=\;\;\argmin_{u^n\in\Xc_b^n}  \;\; \|Au^n-Y^m\|_2^2 \nonumber \\
&\;\;\;\;\;\;{\rm subject \; to} \;\;  \sum\limits_{a^{k+1}\in \Xc_b^{k+1}} w_{a^{k+1}}\hat{p}^{(k+1)}(a^{k+1}|u^n)\leq \g,
\end{align}
or its Lagrangian form 
\begin{align}
  \Xh^n&= \argmin_{u^n\in\Xc_b^n}\Big[\sum_{a^{k+1}\in\Xc_b^{k+1}} w_{a^{k+1}} \hat{p}^{(k+1)}(a^{k+1}|u^n)+{\lambda\over n^2}\|Au^n-Y^m\|^2\Big].\label{eq:Q-MAP-L}
  \end{align}
The choice of  parameters $\lambda>0$  and $\gamma$ is discussed later in our analysis.  We refer to both \eqref{eq:Q-MAP} and \eqref{eq:Q-MAP-L}  as quantized MAP (Q-MAP) estimators.

%
Obtaining the Q-MAP estimator involved  several approximation and relaxation steps.  It is not clear  how accurate these approximations are, and what the performance of the ultimate algorithm is.  Also, solving Q-MAP estimator requires specifying parameters $b$ and $\lambda$, which significantly  affect the performance of the estimator.    These questions are all answered in Section \ref{sec:analysisqmap}. Before that, in the following section,  we  focus on two  specific processes, which are well-studied in the literature,  and  derive the  Q-MAP formulation in each case. This will clarify some of the properties of our Q-MAP formulation.


\section{Special distributions}\label{sec:special-dist}
To get a better understanding of Q-MAP optimization described in \eqref{eq:Q-MAP} and \eqref{eq:Q-MAP-L} and especially the term 
\begin{align}
\sum_{a^{k+1}\in\Xc_b^{k+1}} w_{a^{k+1}} \hat{p}^{(k+1)}(a^{k+1}|u^n),
\end{align} in this section we study two special distributions and derive  simpler statement for  the Q-MAP optimization in each case.

\subsection{Independent and identically distributed sparse processes}\label{sec:iid-sparse}

One of the most popular models for sparse parameter vectors is the spike and slab prior \cite{mitchell1988bayesian}. Consider an i.i.d.~process $\Xbbf$, such that $X_i\sim (1-p)\delta_0+pU_{[0,1]}$.  Since the process is i.i.d., by setting $k=0$ the optimization stated  in \eqref{eq:Q-MAP} can be simplified as 
\begin{align}
  \Xh^n&= \argmin_{u^n\in\Xc_b^n: \sum\limits_{a\in\Xc_b} w_{a} \hat{p}^{(1)}(a|u^n)\leq \g} \|Au^n-Y^m\|^2,\label{eq:QMAP-iid-1}
  \end{align}
  where 
  \begin{align}
  w_a=\log {1\over \P([X_{1}]_b=a_{1})} = \log {1\over (1-p)\ind_{a_1=0}+p{2^{-b}}}.
  \end{align}
Therefore, $\sum_{a\in\Xc_b} w_{a} \hat{p}^{(1)}(a|u^n)$ in \eqref{eq:QMAP-iid-1} can be written as
  \begin{align}
&\sum_{a\in\Xc_b} w_{a} \hat{p}^{(1)}(a|u^n) =\sum_{a\in\Xc_b} w_{a} \Big({1\over n}\sum_{i=1}^n\ind_{u_i=a}\Big)\nonumber\\
&=  {1\over n}\sum_{i=1}^n\sum_{a\in\Xc_b} w_a \ind_{u_i=a} \stackrel{(a)}{=}  {1\over n}\sum_{i=1}^n w_{u_i} \nonumber\\
&=-{1\over n}\sum_{i=1}^n \log ((1-p)\ind_{u_i=0}+p{2^{-b}})\nonumber\\
&=-\hat{p}(0|u^n) \log ((1-p)+p{2^{-b}})-(1-\hat{p}(0|u^n) )\log(p{2^{-b}})\nonumber\\
&=\hat{p}(0|u^n) \log {p{2^{-b}}\over (1-p)+p{2^{-b}}}-\log(p{2^{-b}}),\label{eq:simplified-cost-ell-0}
  \end{align}
  where (a) holds because $\sum_{a\in\Xc_b} w_a \ind_{u_i=a}=w_{u_i}$. Since $-\log(p{2^{-b}})$ is constant, and since 
    \begin{align}
  \a_b\triangleq \log { (1-p)+p{2^{-b}}\over p{2^{-b}}}.\label{eq:alpha-b}
  \end{align}
  is positive, from \eqref{eq:simplified-cost-ell-0}, an upper bound on $\sum_{a\in\Xc_b} w_{a} \hat{p}^{(1)}(a|u^n) $ is in fact an upper bound on the $\ell_0$-norm of $u^n$ defined as
   \begin{align}
  \|u^n\|_0\triangleq |\{i:\; u_i\neq 0\}|.
  \end{align}
(Note that $  \|u^n\|_0=(1-\hat{p}(0|u^n))n$.)  Therefore, given these simplifications, \eqref{eq:QMAP-iid-1} can be written as 
  \begin{align}
  \Xh^n&= \argmin_{u^n\in\Xc_b^n: \|u^n\|_0\leq \g'} \|Au^n-Y^m\|^2 \label{eq:QMAP-iid-2}
  \end{align}
  where $\g'$ is a function of $\g$, $b$ and $p$.
%

\subsection{Piecewise-constant processes} \label{sec:PC}

Another popular example is  the piecewise-constant processes. (Refer to \cite{tibshirani2005sparsity} for some applications of this model.) As our second example, we introduce a first-order Markov process that can model piecewise-constant functions. Conditioned on $X_i=x_i$, $X_{i+1}$ is distributed as $(1-p)\delta_{x_i}+p {\rm Unif}_{[0,1]}$. In other words, at each time step, the Markov chain either stays at its previous value or jumps to a new value, which is drawn from a uniform distribution over $[0,1]$, independent of the past values. The jump process can be modeled as a $\Bern(p)$ process which is independent of the past values of  the Markov chain. Then, since the process has a memory of order one, \eqref{eq:Q-MAP-L} can be written as
\begin{align}
  \Xh^n&= \argmin_{u^n\in\Xc_b^n}\Big[\sum_{a^{2}\in\Xc_b^{2}} w_{a^{2}} \hat{p}^{(2)}(a^{2}|u^n) +{\lambda\over n^2}\|Au^n-Y^m\|^2\Big],\label{eq:QMAP-markov-1}
  \end{align}
  where for $a^2\in\Xc_b^2$
  \begin{align}
  w_{a^2}=\log {1\over \P([X_{2}]_b=a_{2}|[X_{1}]_b=a_{1})}.
  \end{align}
  Given the Kernel of the Markov chain, we have
  \begin{align}
\P([X_{2}]_b=a_{2}|[X_{1}]_b=a_{1})=(1-p)\ind_{a_2=a_1}+p{2^{-b}}.
  \end{align}
  Let $N_J(u^n)$ denote the number of jumps in sequence $u^n$, \ie $N_J(u^n)=\sum_{i=2}^n\ind_{u_i\neq u_{i-1}}$. Then, the first term in the cost function in \eqref{eq:Q-MAP} can be rewritten as 
  \begin{align}
  &\sum_{a^2\in\Xc_b^2} w_{a^2} \hat{p}^{(2)}(a^2|u^n)\nonumber\\
  &=  {1\over n-1}\sum_{i=2}^n\sum_{a^2\in\Xc_b} w_{a^2} \ind_{u_{i-1}^i=a^2}=  {1\over n-1}\sum_{i=2}^n w_{u_{i-1}^i}\nonumber\\
&={1\over n-1}\sum_{i=2}^n \log((1-p)\ind_{u_i=u_{i-1}}+p{2^{-b}})\nonumber\\
&=-{N_J(u^n)\over n-1} \log(p{2^{-b}})-(1-{N_J(u^n)\over n-1}) \log(1-p+p{2^{-b}})\nonumber\\
&={N_J(u^n)\over n-1} \log{1-p+p{2^{-b}} \over p{2^{-b}}}-\log(1-p+p{2^{-b}}).\label{eq:cost-markov}
  \end{align}
  Inserting \eqref{eq:cost-markov} in \eqref{eq:QMAP-markov-1}, it follows that
\begin{align}
  \Xh^n&= \argmin_{u^n\in\Xc_b^n}\Big[\a_b({N_J(u^n)\over n-1})+{\lambda\over n^2}\|Au^n-Y^m\|^2\Big],\label{eq:QMAP-markov-1}
  \end{align}
  where $\a_b$ is defined in \eqref{eq:alpha-b}. Note that the term ${N_J(u^n)\over n-1}$ is counting the number of jumps in $u^n$ which seems to be a natural regularizer here.


\section{Exponential convergence rates}\label{sec:exp-rate}

Consider $X^n$ and its quantized version $[X^n]_b$  generated by the stationary process $\Xbbf$. In  our theoretical analysis,  one of the main features required from the process $\Xbbf$ is that the empirical statistics derived from  $[X^n]_b$ to converge, asymptotically, to their expected values. In all our analysis we require this to hold even when $b$ grows with $n$ slowly enough.  Intuitively speaking, if this is not the case, we do not expect the Q-MAP estimator to be able to obtain a good estimate of $X^n$.  In the following two sections, we study two important classes of stochastic processes which satisfy this property.


\subsection{$\Psi^*$-mixing processes}\label{sec:mixprocess}

The first class of processes that satisfy our requirements are $\Psi^*$-mixing processes. Consider a stationary process $\Xbbf=\{X_i\}_{i=1}^\infty$. Let $\mathcal{F}_j^k$ denote the $\sigma$-field of events generated by random variables $X_j^k$, where $j\leq k$. Define 
\begin{equation}
\psi^*(g) = \sup \frac{\P(\Ac \cap \Bc)}{\P(\Ac) \P(\Bc)},\label{eq:def-Psi-star}
\end{equation}
where the supremum  is taken over all events $\Ac \in \mathcal{F}_{0}^{j}$ and $\Bc \in \mathcal{F}_{j+g}^\infty$, where $P(\Ac)>0$ and $\P(\Bc)>0$. 

\begin{definition} A stationary process $\Xbbf=\{X_n\}_{1}^\infty$ is called $\psi^*$-mixing, if $\psi^*(g) \to 1$, as $g$ grows to infinity.\\
\end{definition}

There are many examples of $\Psi^*$-mixing processes. For instance, it is straightforward to check that any i.i.d.~sequence or any moving average of an i.i.d.~sequence (of finite order) is $\Psi^*$-mixing. Also, every finite-state Markov chain is $\Psi^*$-mixing \cite{Shields}. (For further information on $\Psi^*$-mixing processes the reader may refer to \cite{bradley2005basic}.)   

As mentioned  earlier, the advantage of $\Psi^*$-mixing processes is the fast convergence of their empirical distributions to their expected values. This is captured by the following result from \cite{JalaliP:14-arxiv}, which is a straightforward extension of a similar result in \cite{Shields} for finite-alphabet processes.

\begin{theorem}\label{thm:exp-rate-mixing}
Consider a $\Psi^*$-mixing process $\Xbbf=\{X_i\}_{i=1}^{\infty}$, and its $b$-bit quantized version $\Zbbf=\{Z_i\}_{i=1}^{\infty}$, where $Z_i=[X_i]_b$ and $\Zc=\Xc_b$. Define measure $\mu_k$, such that for $a^k\in\Zc^k$, $\mu_k(a^k)=\P(Z^k=a^k)$. Then, for any $\e>0$, and any $b$ large enough, there exists $g\in\mathds{N}$, depending only on $\e$ and function $\Psi^*(g)$ defined in \eqref{eq:def-Psi-star}, such that for any $n>6(k+g)/\e+k$,
\begin{equation}\label{eq:exponentialconvergence}
\P(\|\hat{p}^{(k)}(\cdot|Z^n)-\mu_k\|_1\geq \e)\leq 2^{c\e^2/8}(k+g)n^{|\Zc|^k}2^{-{nc\e^2\over 8(k+g)}},
\end{equation}
where $c=1/(2\ln 2)$.
\end{theorem}

Note that the upper bound in \eqref{eq:exponentialconvergence} only depends on $b$ through $|\Zc|=|\Xc_b|$. Hence, if $b=b_n$ grows with $n$, it should grow slowly enough, such that overall $2^{c\e^2/8}(k+g)n^{|\Zc|^k}2^{-{nc\e^2\over 8(k+g)}}$ still converges to zero, as $n$ grows to infinity.  One example, which we also use in our results, is $b=b_n=\lceil r \log\log n\rceil$,  $r\geq 1$.  For such choices of $b_n$, Theorem \ref{thm:exp-rate-mixing} guarantees that  the empirical distribution derived from the quantized sequence remains close to its expected value, with high probability.

\subsection{Weak $\Psi^*_q$-mixing  Markov processes}\label{sec:Markov}

Finite-alphabet Markov chains are known to be $\Psi^*$-mixing, and therefore their empirical distributions have exponential convergence rates  \cite{Shields:96}. Continuous space Markov processes on the other hand are not $\Psi^*$-mixing in general and hence we cannot use the results  mentioned in the last section. However, for many of such processes, it is still possible to show that the empirical distribution of their quantized version converges to its expected value, even if the quantization level $b$ grows with $n$, at a slow-enough rate. In this section, we show how this result  can be proved.  


Consider an analog stationary stochastic process $\Xbbf=\{X_i\}$, with alphabet $\Xc=[l,u]$, where $l,u\in\mathds{R}$. Let process $\Zbbf=\{Z_i\}$ denote the $b$-bit quantized version of process $\Xbbf$. That is, $Z_i=[X_i]_b$, and the alphabet of process $\Zbbf$ is $\Zc=\Xc_b=\{[x]_b: x\in\Xc_b\}$. Also, let $\mu_k^{(b)}$ denote the distribution of $Z^k$. That is, for any $z^k\in\Zc^k$,
\begin{align}
\mu_k^{(b)}(z^k)=\P(Z^k=z^k)=\P([X^k]_b=z^k).
\end{align}
The following lemma proves that if the process $\Xbbf$ has a property analogous to being $\Psi^*$-mixing, then potentially it has exponential convergence rates. 

\begin{lemma}\label{lemma:Psi-dist}
Suppose that the stationary process $\Xbbf$ is such that there exists a function $\Psi:\mathds{N}\times \mathds{N}\to\mathds{R}^+$, which satisfies the following. For any $(b,g,\ell_1,\ell_2)\in\mathds{N}^4$, $u^{\ell_1}\in\Zc^{\ell_1}$, and  $v^{\ell_2}\in\Zc^{\ell_2}$:
\begin{align}
\P(Z^{\ell_1}=u^{\ell_1},Z_{\ell_1+g+1}^{\ell_1+g+\ell_2}=v^{\ell_2})\leq \P(Z^{\ell_1}=u^{\ell_1})
\P(Z_{\ell_1+g+1}^{\ell_1+g+\ell_2}=v^{\ell_2})\Psi(b,g),\label{eq:cond-Psi-b-g}
\end{align}
where $b$ denotes the quantization level of process $\Zbbf$.
Then for any given $\e>0$, for any positive integers $k$ and $g$ such that $4(k+g)/(n-k)<\e$, 
\begin{align}
\P(\|\hat{p}^{(k)}(\cdot|Z^n)-\mu_k^{(b)}\|_1\geq \e)\leq (k+g)\Psi^t(b,g)(t+1)^{|\Zc|^k}2^{-c\e^2t/4}.
\end{align}
where $t=\lfloor{n-k+1\over k+g}\rfloor$ and $c=1/(2\ln 2)$.
\end{lemma}

The proof is presented in Section \ref{proof:lemma2}. Note that if the process is $\Psi^*$-mixing then it is straightforward to confirm the existence of $\tilde{\Psi}(g)$ that satisfies
\begin{align}
\P(Z^{\ell_1}=u^{\ell_1},Z_{\ell_1+g+1}^{\ell_1+g+\ell_2}=v^{\ell_2})\leq \P(Z^{\ell_1}=u^{\ell_1})
\P(Z_{\ell_1+g+1}^{\ell_1+g+\ell_2}=v^{\ell_2})\tilde{\Psi}(g).
\end{align}
However, in this section, we are interested in processes that are not necessarily $\Psi^*$-mixing. Lemma \ref{lemma:Psi-dist} allows us to prove the convergence of the empirical distributions for many such  processes. To justify our claims, we focus on the class of Markov processes. For notational simplicity we focus on first order Markov processes. It is straightforward to extend these results to higher  order Markov processes as well.

Let $X$ denote a first-order stationary  Markov process with Kernel function $K: (\Xc,2^{\Xc}) \rightarrow \mathbb{R}^+$ and stationary distribution $\pi: 2^{\Xc}\rightarrow \mathds{R}^+$. (Here $2^{\Xc}$ denotes the set of subsets of $\Xc$.) In other words, for any $x\in\Xc$ and any measurable subset of  $\Xc$,
\begin{align}
K(x,\Ac)=\P(X_2\in\Ac|X_1=x),
\end{align}
and
\begin{align}
\pi(\Ac)=\P(X_1\in\Ac).
\end{align}
Also, for $g\in\mathds{N}^+$, 
\begin{align}
K^g(x,\Ac)=\P(X_{1+g}\in\Ac|X_1=x).
\end{align}
Clearly $k^g$ can be evaluated from function $K$. 
Finally,  with a slight overloading of notation, for $x\in\Xc$ and $z\in\Xc_b$, 
\begin{align}
\pi(z)=\P([X_1]_b=z),
\end{align}
and
\begin{align}
K(x,z)=\P([X_2]_b=z|X_1=x),
\end{align}
Similarly, for $g\in\mathds{N}^+$, $K^g(x,z)=\P([X_{1+g}]_b=z|X_1=x)$.  Again, with another slight overloading of notation, for $x\in\Xc$, and $w^{l+1}\in\Zc^{l+1}$,  
\begin{align}
\pi(w_2^{l+1}|x)={\P([X_2^{l+1}]_b=w^{l+1}_2|X_1=x)},
\end{align}
and
\begin{align}
\pi(w^l|w_0)={\P([X_2^{l+1}]_b=w_2^{l+1}|[X_1]_b=w_1)}.
\end{align}

Define the functions $\Psi_1:\mathds{N}\times\mathds{N}\to \mathds{R}^+$ and $\Psi_2:\mathds{N}\to \mathds{R}^+$ as 
\begin{align}
\Psi_1(b,g) \triangleq \sup_{(x,z)\in\Xc\times\Xc_b} {K^{g}(x,z) \over \pi(z)},\label{eq:Psi1-def}
\end{align}
and
\begin{align}
\Psi_2(b)\triangleq \sup_{(x,\ell_2,w^{\ell_2})\in\Xc\times\mathds{N}\times \Zc^{\ell_2}: [x]_b=w_1} {\pi(w_{2}^{\ell_2}|x)\over \pi(w_{2}^{\ell_2}|w_1)}.\label{eq:Psi2-def}
\end{align}

Our next lemma shows how the function $\Psi(b,g)$ in Lemma \ref{lemma:Psi-dist} can be calculated from the functions $\Psi_1$ and $\Psi_2$.

\begin{lemma}\label{lemma:Psi-Markov}
Consider a first-order aperiodic Markov process $\Xbbf=\{X_i\}_{i=1}^{\infty}$. Let $\Zbbf=\{Z_i\}$ denote the $b$-bit quantized version of process $\Xbbf$. That is, $Z_i=[X_i]_b$, and $\Zc=\Xc_b=\{[x]_b: x\in\Xc\}$. Also, let $\mu_b$ denote the distribution  associated with the finite-alphabet  process $\Zbbf$. Then, for all $(\ell_1,g,\ell_2)\in\mathds{N}^3$, $u^{\ell_1}\in\Zc^{\ell_1}$, $v^g\in\Zc^g$ and $w^{\ell_2}\in\Zc^{\ell_2}$, we have
 \begin{align}
 \mu_b(u^{\ell_1}v^{g}w^{\ell_2})\leq  \mu_b(u^{\ell_1})\mu_b(w^{\ell_2})\Psi_1(b,g)\Psi_2(b),
 \end{align}
 where by definition  $\mu_b(u^{\ell_1}v^{g}w^{\ell_2})=\P(Z^{\ell_1+g+\ell_2}=[u^{\ell_1},v^{g},w^{\ell_2}])$, 
  $\mu_b(u^{\ell_1})=\P(Z^{\ell_1}=u^{\ell_1})$
  and 
   $\mu_b(w^{\ell_2})=\P(Z^{\ell_2}=w^{\ell_2})$. Furthermore, $\Psi_1(b,g)$ is non-increasing function of $g$ that converges to $1$, for any fixed $b$, as $g$ grows to infinity.
\end{lemma}
The proof is presented in Section \ref{proof:lemma3}.

Note that if we combine Lemmas \ref{lemma:Psi-dist} and \ref{lemma:Psi-Markov} we obtain an upper bound of the form $(k+g)\Psi^t(b,g)(t+1)^{|\Zc|^k}2^{-c\e^2t/4}$ for $\P(\|\hat{p}^{(k)}(\cdot|Z^n)-\mu_k^{(b)}\|_1\geq \e)$. To prove our desired convergence results, we need to ensure that the upper bound on this probability goes to zero as $n \rightarrow \infty$. It is straightforward to note that as $n \rightarrow \infty$, $t \rightarrow \infty$ and hence the term $2^{-c\e^2t/4}$ converges to zero. However, if $\Psi^t(b,g)(t+1)^{|\Zc|^k}$ grows faster than $2^{-c\e^2t/4}$, then we do not reach the desired goal. Our next theorems prove that under some mild conditions on the Markov process, for slow enough growth of $b=b_n$ with $n$, $\Psi^t(b,g)$ does not grow very fast. First note that since for any $z^n\in\Zc^n$, $\| \mu_{k_1}-\hat{p}^{(k_1)}(\cdot|z^n)\|_1\leq \| \mu_{k_2}-\hat{p}^{(k_2)}(\cdot|z^n)\|_1$, for all $k_1\leq k_2$, to prove fast convergence of $\hat{p}^{(k)}(\cdot|Z^n)$ it is enough to prove this statement  for $k$ large.

\begin{theorem}\label{thm:exp-rate-markov}
Consider  an analog aperiodic stationary first-order Markov chain  $\Xbbf$, and its $b$-bit quantized version $\Zbbf$, where $Z_i=[X_i]_b$  and $\Zc=\Xc_b$. Let $\mu_k^{(b)}$ denote the $k$-th order probability distribution of process $\Zbbf$, \ie for any $z^k\in\Zc^k$,
\begin{align}
\mu_k^{(b)}(z^k)=\P(Z^k=z^k).
\end{align}
Let $b=b_n= \lceil r\log\log n \rceil$, where $r\geq 1$. Assume that there exists a sequence $g=g_n$, such that $g_n=o(n)$, and process $\Xbbf$ satisfies the following conditions:
\begin{enumerate}
\item $\lim_{n\to\infty}\Psi_1(b_n,g_n)=1 $, and
\item $\lim_{b\to\infty}\Psi_2(b)=1$,
\end{enumerate}
where functions $\Psi_1$ and $\Psi_2$ are defined in \eqref{eq:Psi1-def} and \eqref{eq:Psi2-def}, respectively. 
Then, given $\e>0$ and positive integer $k$, for $n$ large enough, 
\begin{align}
\P( \|\hat{p}^{(k)}(\cdot|Z^n)-\mu_k^{(b)}\|_1\geq \e)\leq  2^{c\e^2/4} (k+g_n)n^{|\Zc|^k}2^{-{c \e^2n\over 8(k+g_n)}},
\end{align}
where $c=1/(2\ln 2)$. 
\end{theorem}
The proof is presented in Section \ref{proof:thm6}.

\begin{remark}
Lemma \ref{lemma:Psi-Markov} proves that for any fixed $b$, $\Psi_1(b,g)$ converges to one, as $g$ grows without bound. However, in this paper we are mainly interested in the cases where  $ b_n=\lceil r\log\log n \rceil$. The condition on $\Psi_1$ specified in  Theorem \ref{thm:exp-rate-markov} ensures that even if $b$ also grows to infinity, there exists a proper choice of sequence $g_n$ as a function of  $n$, for $\Psi_1(b_n,g_n)$ still converges to one, as $n$ grows without bound. 
\end{remark}

Theorem \ref{thm:exp-rate-markov} proves that if $\lim_{n\to\infty}\Psi_1(b_n,g_n)=1 $ and $\lim_{b\to\infty}\Psi_2(b)=1$, then the quantized version of an analog Markov process also has fast convergence rates. We refer to a Markov process that satisfies these two conditions with $b_n = \lceil r\log\log n \rceil$ as a {\em weak $\Psi^*_q$-mixing Markov} process. 
To better understand these conditions, we next consider the piecewise-constant source studied in Section \ref{sec:PC} and prove that it is a weak $\Psi^*_q$-mixing Markov process.

\begin{theorem}\label{thm:Markov-Psi1-Psi2}
Consider a first-order stationary Markov process  $\Xbbf$,  such that conditioned on $X_i=x_i$, $X_{i+1}$ is distributed as $(1-p)\delta_{x_i}+pf_c$, where $f_c$ denotes an absolutely continuous distribution over $\Xc=[0,1]$. Further assume that there exists $f_{\min}>0$, such that $f_c(x)\geq f_{\min},$ 
for all $x\in\Xc$. Then,
\begin{enumerate}
\item for $b=b_n=\lceil r\log\log n\rceil $ and $g=g_n=\lfloor \g r\log\log n \rfloor$,  where   $\g>-{1\over \log(1-p)}$, $\lim_{n\to\infty}\Psi_1(b_n,g_n)=1$, and
\item $\Psi_2(b)=1$, for all $b$.
\end{enumerate}
\end{theorem}
The proof is presented in Section \ref{proof:thm7}.


 \section{Theoretical analysis of Q-MAP} \label{sec:analysisqmap}

 In order to recover $X^n$ from $m<n$ response variables, intuitively, the process should be of ``low-complexity''. Hence, before performing  the  theoretical  analysis of the Q-MAP optimization, in Section \ref{sec:lowcomplex}, we briefly review a measure of  complexity developed for continuous-valued stochastic processes. This measure plays a pivotal role in our analysis of Q-MAP.   
  \subsection{Low-complexity stochastic processes}\label{sec:lowcomplex}

Consider a stationary  process $\Xbbf=\{X_i\}_{i=1}^{\infty}$ and define its $b$-bit quantized version as  $[\Xbbf]_b=\{[X_i]_b\}_{i=1}^{\infty}$. Since $[\Xbbf]_b$ is derived from a stationary coding of a stationary  process, it is also stationary.  In \eqref{eq:first_appearanced_k} we defined the $k$-th order upper information dimension of a process $\Xbbf$ as
\begin{align}
\bar{d}_k(\Xbbf)=\limsup_{b\to \infty} {H([X_{k+1}]_b|[X^k]_b) \over b}.
\end{align}
Similarly, the $k$-th order lower information dimension of  $\Xbbf$ is defined as $\underline{d}_k(\Xbbf)=\liminf_{b\to \infty} {H([X_{k+1}]_b|[X^k]_b) \over b}.$

\begin{definition}[Upper/lower information dimension \cite{JalaliP:14-arxiv}]
For a stationary  process $\Xbbf$, if $\lim_{k\to\infty}\bar{d}_k(\Xbbf)$ exists, we define the upper information dimension of process $\Xbbf$ as
\begin{align}
\bar{d}_o(\Xbbf)=\lim_{k\to\infty}\bar{d}_k(\Xbbf).
\end{align}
Similarly, if  $\lim_{k\to\infty}\underline{d}_k(\Xbbf)$ exists, the  lower information dimension of process $\Xbbf$ is defined as $\underline{d}_o(\Xbbf)=\lim_{k\to\infty}\underline{d}_k(\Xbbf).$  If $\underline{d}_o(\Xbbf) =\bar{d}_o(\Xbbf)$, ${d}_o(\Xbbf)\triangleq \underline{d}_o(\Xbbf) =\bar{d}_o(\Xbbf)$ is defined as the information dimension of the process $\Xbbf$.\\
\end{definition}
The information dimension of a stationary stochastic process is an extension of the R\'enyi's notion of information dimension defined for random variables and random vectors \cite{Renyi:59}.  As argued in \cite{JalaliP:14-arxiv}, ${d}_o(\Xbbf)$ serves as a measure of complexity for  stochastic process $\Xbbf$ and is related to the number of response variables required for its accurate recovery. Also, as long as $H(\lfloor X_1 \rfloor)<\infty$, ${d}_o(\Xbbf)\leq 1$.  Therefore, if a process is ``structured'' or of ``low-complexity'', ${d}_o(\Xbbf)$ is expected to be strictly smaller than one.

\subsection{Performance of Q-MAP}\label{sec:noiselesstheory}

In this section, we formalize Informal Result 1 presented in the introduction. The following theorem provides conditions for the success of the Q-MAP estimator, for the case where the response variables are noise-free. We state all the results for $\Psi^*$-mixing processes, but they also apply to weak $\Psi^*_q$-mixing Markov processes. 

\begin{theorem}\label{thm:mainresult-Q-MAP}
 Consider $X^n$ generated by a $\Psi^*$-mixing stationary process $\Xbbf$, and let $Y^m=AX^n$. Assume that the entries of  the design matrix $A$ are i.i.d.~$\Nc(0,1)$. Choose  $k$, $r>1$ and $\d>0$, and let $b=b_n=\lceil r\log\log n\rceil$,  $\g=\g_n= b_n(\bar{d}_k(\Xbbf)+\d$), and  $m=m_n\geq (1+\delta)n\bar{d}_k(\Xbbf)$. Assume that  there exists a constant  $f_{k+1}>0$, such that  for  any measurable set  $\Sc_{k+1}\in\Xc^{k+1}$ with $\mu(\Sc_{k+1})>0$, 
\begin{align}
\P(X^{k+1}\in \Sc_{k+1})\geq f_{k+1} |\Sc_{k+1}|,
\end{align}
where $|\Sc_{k+1}|$ denotes the Lebesgue measure of set $\Sc_{k+1}$ and $X^{k+1}$ is generated by source $\Xbbf$.  Further, assume that $\Xh^n$ denotes the solution of  \eqref{eq:Q-MAP}, where the coefficients are computed according to \eqref{eq:coeffs-Q-MAP}. Then, for any $\e>0$,
 \begin{align}
\lim_{n\to\infty} \P( { 1\over \sqrt{n}}\|X^n-\Xh^n\|_2>\e)=0.
  \end{align}
\end{theorem} 
The proof is presented in Section \ref{sec:proof-thm2}.

We remind the reader that we also introduced a Lagrangian version of Q-MAP in \eqref{eq:Q-MAP-L}. It turns out that we can derive the same performance guarantees for the Lagrangian Q-MAP as well. 

\begin{theorem}\label{thm:mainresult-Q-MAP-L}
Consider  a $\Psi^*$-mixing stationery  process $\Xbbf$.  Let $Y^m=AX^n$, where the entries of $A$ are i.i.d. $\Nc(0,1)$. Choose  $k$, $r>1$ and $\d>0$, and let $b=b_n=\lceil r\log\log n\rceil$,  $\l=\l_n=(\log n)^{2r}$ and  $m=m_n\geq (1+\delta)n\bar{d}_k(\Xbbf)$. Assume that  there exists a constant  $f_{k+1}>0$, such that  for  any measurable set  $\Sc_{k+1}\in\Xc^{k+1}$ with $\P(X^{k+1}\in\Sc_{k+1})>0$, 
\begin{align}
\P(X^{k+1}\in\Sc_{k+1})\geq f_{k+1} |\Sc_{k+1}|,
\end{align}
where $|\Sc_{k+1}|$ denotes the Lebesgue measure of set $\Sc_{k+1}$.  Further, assume that $\Xh^n$ denotes the solution of  \eqref{eq:Q-MAP-L}, where the coefficients are computed according to \eqref{eq:coeffs-Q-MAP}. Then, for any $\e>0$,
 \begin{align}
\lim_{n\to\infty} \P( { 1\over \sqrt{n}}\|X^n-\Xh^n\|_2>\e)=0.
  \end{align}
\end{theorem} 
The proof is presented in Section \ref{sec:proof-thm3}.

To better understand the implications of Theorems \ref{thm:mainresult-Q-MAP} and \ref{thm:mainresult-Q-MAP-L},  consider the case where the process is stationary and memoryless. All such processes are $\Psi^*$-mixing, and satisfy   $\bar{d}_k(\Xbbf)=\bar{d}_0(\Xbbf)$, for all $k\geq 0$.   Therefore, as long as $m_n \geq (1+\delta)n\bar{d}_0(\Xbbf)$, asymptotically, the Q-MAP algorithm provides an accurate estimate of the parameter vector. On the other hand,  since the process is i.i.d., $\bar{d}_0(\Xbbf)=\bar{d}(X_1)$, where $\bar{d}(X_1)$ denotes  the upper R\'enyi information dimension  of $X_1$ \cite{Renyi:59}.  On the other hand, for an i.i.d.~process whose  marginal distribution is   a mixture of discrete and continuous distributions, asymptotically, the R\'enyi information dimension of the marginal distribution  characterzies  the minumum number of response variables normalized by $n$ from which accurate recovery of the parameter vector is still possible \cite{WuV:10}. Hence, for such i.i.d.~sources, in a noiseless setting,  the algorithm presented in   \eqref{eq:QMAP-iid-2} achieves the fundamental limits.

Finally, another interesting implication of Theorem \ref{thm:mainresult-Q-MAP-L} is the following. The Q-MAP optimization mentioned in \eqref{eq:Q-MAP} is not a convex optimization. Hence, its solution does not necessarily coincide with the solution of \eqref{eq:Q-MAP-L}. However, at least in the noiseless setting we can derive similar performance bounds for both.

\section{Solving Q-MAP}\label{sec:solving-QMAP}

\subsection{Projected Gradient Descent (PGD)}

The goal of this section is to analyze the performance of the projected gradient descent (PGD) algorithm  introduced in Section \ref{ssec:contributions}.The results are presented for $\Psi^*$-mixing processes, but they also apply to weak $\Psi_q^*$-mixing Markov processes wit no change. 
Note that even though PGD algorithms have been studied extensively for convex optimization problems, since our optimization is discrete and consequently not convex, those analyses do not apply to our problem. Given $\d>0$, consider the Q-MAP optimization characterized as
\begin{align}
\Xh^n=&\argmin_{u^n\in\Xc_b^n}\;\;\;\;\;\;\;\; \|Y^m-Au^n\|^2  \nonumber\\
& {\rm subject \; to}\; \sum_{a^{k+1}\in\Xc_b^{k+1}} w_{a^{k+1}}\hat{p}^{(k+1)}(a^{k+1}|u^n) \leq b(\bar{d}_k(\Xbbf)+\d). 
\end{align}
The corresponding PGD algorithm proceeds as follows. For $t=1,2,\ldots,t_n$,
\begin{align}
S^n(t+1)&=\Xh^n(t)+\mu A^T(Y^m-A\Xh^n(t))\nonumber\\
\Xh^n(t+1)&=\argmin_{u^n\in\Fc_o}\left\|u^n-S^n(t+1)\right\|,\label{eq:PGD-update}
\end{align}
where
\begin{align}
\Fc_o \triangleq \Big\{u^n\in\Xc_b^n: \;  \sum_{a^{k+1}\in\Xc_b^{k+1}} w_{a^{k+1}}\hat{p}^{(k+1)}(a^{k+1}|u^n) \leq b(\bar{d}_k(\Xbbf)+\d) \Big\}.\label{eq:def-Fc-rand}
\end{align}

Theorem \ref{thm:PGD-analysis-noisy} below  proves that, having enough number of noiseless response variables, with probability approaching one,  the PGD-based algorithm recovers  parameters $X^n$ with arbitrarily small error.


So far we have assumed that there is no noise in the response variables. Of course in reality noise is always present. The next result shows that even in the presence of noise, with a higher  number of measurements, the PGD method is still able to recover the signal. 

\begin{theorem}\label{thm:PGD-analysis-noisy}
Consider a $\Psi^{*}$-mixing process $\Xbbf$ and $X^n$ generated by process $\Xbbf$.  Assume that  there exists a constant  $f_{k+1}>0$, such that  for  any measurable set  $\Sc_{k+1}\in\Xc^{k+1}$ with $\P(X^{k+1}\in\Sc_{k+1})>0$, $\P(X^{k+1}\in\Sc_{k+1})\geq f_{k+1} |\Sc_{k+1}|,$
where $|\Sc_{k+1}|$ denotes the Lebesgue measure of set $\Sc_{k+1}$.  Let $Y^m=AX^n+Z^m$, where the elements of matrix $A$ are drawn standard normal  and $Z_i$, $i=1,\ldots,m$, are i.i.d.~$\Nc(0,\sigma^2)$ . For $r>1$, let  $b=b_n=\lceil r\log\log n\rceil,$
and $m=m_n=80 nb(\bar{d}_k(\Xbbf)+\d).$ Let $\mu={1\over m}$, and consider  $\Xh_n(t)$, $t=0,1,\ldots,t_n$, generated according to \eqref{eq:PGD-update}. Define the error vector at iteration $t$ as \begin{align}E^n(t)=\Xh^n(t)-[X^n]_b.\end{align}
 Then, with probability approaching one,
\begin{align}
{1\over \sqrt{n}}|E^n(t+1)\|\leq {0.9\over \sqrt{n}}\|E^n(t)\|+2\Big(2+\sqrt{n\over m}\;\Big)^2 2^{-b}+ { \sigma\over 2}\sqrt{b(\bar{d}_k(\Xbbf)+3\d)\over m},
\end{align}
for $t=1,2,\ldots$.

\end{theorem}

Comparing  this result with Theorem \ref{thm:mainresult-Q-MAP-L} reveals  that the minimum value of $m$ required  in this theorem is $80b_n = 80\lceil r\log\log n\rceil $ times higher than the number of response variables required in Theorem \ref{thm:mainresult-Q-MAP-L}. One can also decrease(increase) the factor $10$ and slow down (speed up) the convergence rate of the algorithm. At this point it is not clear to us whether the factor $r\log\log n$ in Theorem \ref{thm:PGD-analysis-noisy} (for the number of response variables) is necessary in general or is an artifact of our proof techniques. For some specific priors such as the the spike and slab distribution discussed earlier,  with a slight modification of the algorithm, it is known that this factor can be improved. In that case, given the special form of the coefficients, we may let $b$ grow to infinity for a fixed $n$. Then the algorithm becomes equivalent to the iterative hard thresholding (IHT) algorithm introduced in \cite{BlumensathD:09}. The analysis in \cite{BlumensathD:09} shows that the number of response variables $m_n$ required by the  IHT algorithm is proportional to $n$ and does not have the $\log \log n$ factor that appears in Theorem \ref{thm:PGD-analysis-noisy}.

In Theorem \ref{thm:PGD-analysis-noisy} and all of the previous results,  the elements of the design matrix $A$ were assumed to be generated according to $\Nc(0,1)$ distribution. In the noisy setup, where the response variables are distorted by a noise of variance $\sigma^2$,  this model implies having per response signal to noise ratio (SNR) that grows linearly with $n$. This has made the result of the previous theorem misleading. If we consider per element error, i.e., $\frac{\|E^n(t+1)\|}{\sqrt{n}}$ then the error seems to go to zero. To fix this issue, we assume that the elements of $A$ are generated according to $\Nc(0,{1\over n})$.  The following corollary restates the result of Theorem \ref{thm:PGD-analysis-noisy} under this scaling and an appropriate adjustment of coefficient $\mu$.

\begin{corollary}
Consider the setup of Theorem \ref{thm:PGD-analysis-noisy}, where the elements of $A$ are generated i.i.d.\ $\Nc(0,{1\over n})$.
Let 
\begin{align}
S^n(t+1)&=\Xh^n(t)+{n\over m}A^T(Y^m-A\Xh^n(t))\nonumber\\
\Xh^n(t+1)&=\argmin_{u^n\in\Fc_o}\left\|u^n-S^n(t+1)\right\|.\label{eq:PGD-noisy-update}
\end{align}
 Then, with probability approaching one,
\begin{align}
{1\over \sqrt{n}}\|E^n(t+1)\|\leq {0.9\over \sqrt{n}}\|E^n(t)\|+{2(\sqrt{n}+2\sqrt{m})^2\over m}2^{-b}+ { \sigma\over 2}\sqrt{nb(\bar{d}_k(\Xbbf)+3\d)\over m}.
\end{align}
for $t=1,2,\ldots,t_n$.
\end{corollary}
Note that for  $m=m_n=80 nb(\bar{d}_k(\Xbbf)+3\d)$, ${ \sigma\over 2}\sqrt{nb(\bar{d}_k(\Xbbf)+3\d)\over m}\leq {\sigma\over 12}$.

\subsection{Discussion of Computational complexity of PGD}\label{sec:comp-comp}

As explained earlier, at each iteration,  the PGD-based algorithm updates its estimate  $\Xh^n(t )$ to $\Xh^n(t +1)$ by performing the following  two steps:
\begin{enumerate}
\item $S^n(t+1)=\Xh^n(t)+\mu A^T(Y^m-A\Xh^n(t))$,
\item $\Xh^n(t+1)=\argmin_{u^n\in\Fc_o}\left\|u^n-S^n(t+1)\right\|$.
\end{enumerate}
Clearly the challenging part is performing the second step, which is projection on the set $\Fc_o$. For some special distributions, such as the spike and slab prior, discussed in Section \ref{sec:iid-sparse}, and piecewise-constant processes, discussed in Section \ref{sec:PC}, and their extensions this projection step is not complicated. For instance, for the aforementioned sparse vector, $\Fc_o$ contains sparse quantized vectors, and hence the projection step is just keeping the quantized versions of the largest components of $S^{t}$ and setting the rest to zero. This is very similar to the IHT algorithm  \cite{BlumensathD:09}. However, for more general distributions this projection step may be challenging.  

Hence, in order to make the PGD method efficient, we need to be able to solve the following optimization efficiently: 
\begin{align}\label{eq:1}
\xh^n\;=\;&\argmin_{u^n\in\Xc_b^{n}} \;\;\;\;\;\;\;\left\|u^n-x^n\right\| \nonumber \\
&{\rm subject \; to} \;\; \;\sum_{a^{k+1}\in\Xc_b^{k+1}} w_{a^{k+1}}\hat{p}^{(k+1)}(a^{k+1}|u^n) \leq \gamma
\end{align}
where $x^n\in\mathds{R}^n$,  weights $\{w_{a^{k+1}}: a^{k+1}\in \Xc_b^{k+1}\}$ and $\gamma\in\mathds{R}^+$ are all given input parameters. 
  Equation \eqref{eq:1} can be stated in the  Lagrangian form as
\begin{align}
\xh^n\;=\;&\argmin_{u^n\in\Xc_b^{n}} \Big[{1\over n^2}\left\|u^n-x^n\right\|^2 +\a \sum_{a^{k+1}\in\Xc_b^{k+1}} w_{a^{k+1}}\hat{p}^{(k+1)}(a^{k+1}|u^n)\Big],\label{eq:lagrangian-eq-projection}
\end{align}
where $\a>0$ is a parameter that depends on $\gamma$.
Since $\|u^n-x^n\|^2=\sum_{i=1}^n(u_i-x_i)^2$, the optimization stated in \eqref{eq:lagrangian-eq-projection} is exactly the optimization studied  in \cite{JalaliM:12}. It has been proved in \cite{JalaliM:12}  that the solution of \eqref{eq:lagrangian-eq-projection} can be found efficiently via the standard dynamic programming (Viterbi algorithm) \cite{viterbi1967error}.For further information you may refer to  \cite{JalaliM:12}. 

The question is whether, for an appropriate choice of $\a$,  the minimizers of \eqref{eq:1} and \eqref{eq:lagrangian-eq-projection} are the same. If the answer to this question is affirmative, it implies that both steps of the PGD method can be implemented efficiently. In the following we intuitively argue why we believe   this might be the case. Making the argument rigorous and a deeper investigation of this connection is left to future research. 

Consider partitioning the set of sequences in $\Xc_b^n$ based on their $(k+1)$-th order empirical distributions, which are referred to as their  $(k+1)$-types.  For any possible  $(k+1)$-type $q_{k+1}(\cdot): \Xc_b^{k+1}\to \mathds{R}^+$, let $\Tc_n(q_{k+1})$ denote the set of sequences in $\Xc_b^n$ whose  $(k+1)$-types agree with  $q_{k+1}$. That is,
\begin{align}
\Tc_n(q_{k+1})\triangleq \{u^n: \hat{p}^{(k+1)}(a^{k+1}|u^n)=q_{k+1}(a^{k+1}), \forall \; a^{k+1}\in\Uc_b^{k+1}\}.
\end{align}
 Let $\Pc_{n}^{k+1}$ denote the set of all possible $(k+1)$-types, for sequences in $\Xc_b^n$. In other words,
\begin{align}
\Pc_{n}^{k+1}\triangleq \{  \hat{p}^{(k+1)}(\cdot|u^n): u^n\in\Xc_b^n\}.
\end{align}
 It can be proved that (Theorem I.6.14 in \cite{Shields:96})
\begin{align}
|\Pc_{n}^{k+1}|\leq (n+1)^{|\Xc_b|^{k+1}}.
\end{align}
Furthermore, we have
\begin{align}
\Xc_b^n=\cup_{q_{k+1}\in\Pc_{n}^{k+1}} \Tc_n(q_{k+1}).
\end{align}
Therefore,  
\begin{align}
&\min_{u^n\in\Xc_b^{n}} \Big[{1\over n}\left\|u^n-x^n\right\|^2 +\a \sum_{a^{k+1}\in\Xc_b^{k+1}} w_{a^{k+1}}\hat{p}^{(k+1)}(a^{k+1}|u^n)\Big]\nonumber\\
&\;=\;\min_{q_{k+1} \in \Pc_n^{k+1}}\min_{u^n\in\Tc_n(q_{k+1}) }   \Big[{1\over n}\left\|u^n-x^n\right\|^2 +\a \sum_{a^{k+1}\in\Xc_b^{k+1}} w_{a^{k+1}}q_{k+1}(a^{k+1}\Big]\nonumber\\
&\;=\;\min_{q_{k+1} \in \Pc_n^{k+1}} \left[ \Big [\min_{u^n\in\Tc_n(q_{k+1}) }   {1\over n}\left\|u^n-x^n\right\|^2 \Big]+\a \sum_{a^{k+1}\in\Xc_b^{k+1}} w_{a^{k+1}}q_{k+1}(a^{k+1})\right],\label{eq:2}
\end{align}
where the last  line follows because $ \sum_{a^{k+1}\in\Xc_b^{k+1}} w_{a^{k+1}}\hat{p}^{(k+1)}(a^{k+1}|u^n)$ only depends on the $k$-type of sequence $u^n$.
For any type  $q_{k+1} \in \Pc_n^{k+1}$ define the minimum distortion attainable by sequences of that type as $D(q_{k+1},x^n)$, \ie
\begin{align}
D(q_{k+1},x^n) = \min_{u^n \in \Tc_n(q_{k+1}) }{1\over n} \left\|u^n-x^n\right\|^2,
\end{align}
Then \eqref{eq:2} and  \eqref{eq:1} can be written as
\begin{align}
\min_{q_{k+1} \in \Pc_{n}^{k+1}} \Big[D(q_{k+1},x^n) +\a\sum_{a^{k+1}\in\Xc_b^{k+1}} w_{a^{k+1}}q_{k+1}(a^{k+1}) \Big] 
\end{align}
and
\begin{align}
\min_{q_{k+1} \in \Pc_{n}^{k+1}:  \sum_{a^{k+1}\in\Xc_b^{k+1}} w_{a^{k+1}}q_{k+1}(a^{k+1})\leq \g} D(q_{k+1},x^n),
\end{align}
respectively.  Both of these optimizations are discrete optimization. However, since $ \sum_{a^{k+1}\in\Xc_b^{k+1}} w_{a^{k+1}}q_{k+1}(a^{k+1})$ is a convex function of $q_{k+1}$, if, in the high-dimensional setting, for  input sequences $x^n$ of interest,  $D(q_{k+1})$ also behaves almost as a convex function, then we expect  the two optimizations to be the same, for a proper choice of parameter $\a$.  In the remainder of this section, we argue why,  in a high dimensional setting, we conjecture that  $D(q_{k+1})$  satisfies the mentioned property. We leave further investigation of the subject to future research. 
%

First note that if $x^n$ is almost stationary, for instance it is generated by a Markov process with finite memory,  then $D(q_{k+1},x^n)$ only depends on $q_{k+1}$ and  finite-order empirical distributions of $x^n$, and not on $n$ or $x^n$. Now assuming that this is true,  consider $q^{(1)}_{k+1}$ and $q^{(2)}_{k+1}$ in $\Pc_n^{k+1}$. Also given $\theta\in(0,1)$, let $n_1=\lfloor \theta n\rfloor$ and $n_2=n-n_1$. Also, let  $\xt^{n_1}$ and $\bar{x}^{n_2}$ denote the minimizers of $D(q^{(1)}_{k+1},x^{n_1})$ and $D(q^{(2)}_{k+1},x_{n_1+1}^{n})$, respectively.  Assume that $\theta q^{(1)}_{k+1}+(1-\theta)q^{(2)}_{k+1}\in \Pc_n^{k+1}$ and let 
$\xh^n=[\xt^{n_1},\bar{x}^{n_2}]$. Then, for large $n$, it is straightforward to check that $\hat{p}^{(k+1)}(\cdot|\xh^n)\approx \theta \hat{p}^{(k+1)}(\cdot|\xt^{n_1})+(1-\theta)  \hat{p}^{(k+1)}(\cdot|\bar{x}^{n_2})=\theta q^{(1)}_{k+1}+(1-\theta)q^{(2)}_{k+1}$.
\begin{align}
n D(\theta p_1 + (1-\theta)p_2,x^n) &\leq \|x^n-\xh^n\|^2\nonumber\\
&= \|x^{n_1}-\xt^{n_1}\|^2+\|x_{n_1+1}^{n}-\bar{x}^{n_2}\|^2\nonumber\\
&= n_1D(q^{(2)}_{k+1},x_{n_1+1}^{n})+n_2D(q^{(2)}_{k+1},x_{n_1+1}^{n}).
\end{align}
Dividing both sides by $n$, it follows that
\begin{align}
D(\alpha p_1 + (1-\alpha)p_2,x^n)\leq \theta D(q^{(2)}_{k+1},x_{n_1+1}^{n})+(1-\theta)D(q^{(2)}_{k+1},x_{n_1+1}^{n}). \label{eq:3}
\end{align}
Therefore,   as we expect, if for large values of $n$ and stationary sequences $x^n$, $D(q_{k+1},x^n)$  depends on  $x^n$ only through its  empirical distribution, then in \eqref{eq:3},   since $x_{n_1+1}^{n}$ and $x_{n_1+1}^{n}$ have almost the same empirical distribution as $x^n$, they can be replaced by $x^n$.  This establishes our conjecture about almost convexity of function $D$. 

\section{Proofs}\label{sec:proof}

\subsection{Preliminaries on information theory}

Before presenting the proofs, in this section, we review  some preliminary  definitions and concepts that  are used in some of the proofs. 

Consider stationary process $\Ubbf=\{U_i\}_{i=1}^{\infty}$, with  finite alphabet $\Uc$. The entropy rate of process $\Ubbf$ is defined as
\begin{align}
\bar{H}(\Ubbf)\triangleq \lim_{k\to\infty}H(U_{k+1}|U^k).
\end{align}
Consider  $u^n\in\Uc^n$, where $\Uc$ is a finite set. The  $(k+1)$-th order empirical distribution of $u^n$ is defined in \eqref{eq:emp-dist}. The $k$-th order conditional empirical entropy of $u^n$ is defined as $\hat{H}_k(u^n)=H(U_{k+1}|U^k)$, where $U^{k+1}$ is distributed as $\hat{p}^{(k+1)}(\cdot|u^n)$. In other words,
\begin{align}
\hat{H}_k(u^n)=-\sum_{a^{k+1}\in\Uc^{k+1}}\hat{p}^{(k+1)}(a^{k+1}|u^n)\log{\hat{p}^{(k+1)}(a^{k+1}|u^n) \over \hat{p}^{(k)}(a^k|u^n)}.
\end{align}

In some of the proofs we employ a compression scheme called {\em Lempel-Ziv}. Compression schemes aim to represent a sequence $u^n \in\Uc^n$ in as few bits as possible. It turns out that intuitively speaking if $u^n$ is a sample of a finite-alphabet stationary ergodic process $\Ubbf$, asympotocially, the smallest expected number of  bits per symbol required to represent $u^n$ is  $ \bar{H}(\Ubbf)$. Compression algorithms that achieve this bound are called optimal. One of the well-known examples of optimal compression schemes is Lempel-Ziv (LZ) \cite{LZ} coding. (LZ is also a universal compression code, since it does not use any information regarding the distribution of the process.)

In summary, the LZ compression code operates at follows: it first incrementally parses the input sequence into unique phrases such that each phrase is the shortest phrase that is not seen earlier. Then, each phrase is encoded by i) an index to the location of the phrase consisting of the current phrase except its last symbol, and ii) last symbol of the phrase. 


 Given $u^n\in\Uc^n$, let  $\ell_{\rm LZ}(u^n)$ denote the length of the binary coded sequence assigned to $u^n$ using the LZ compression code. Note that since LZ algorithm assigns a unique coded sequence to every input sequence, we have 
 \begin{align}
|\{u^n: \ell_{\rm LZ}(u^n)\leq r\}|\leq \sum_{i=1}^r2^i\leq 2^{r+1}.\label{eq:LZ-sequences}
\end{align}
The LZ length function $\ell_{\rm LZ}(\cdot)$ is mentioned in some of the following proofs because of its connections with the conditional empirical entropy function $\hat{H}_k(\cdot)$.  This connection established   in \cite{PlotnikW:92} for binary sources and extended in \cite{JalaliP:14-arxiv} to general sources with alphabet $\Uc$ such that $|\Uc|=2^b$ states that, for all $u^n\in\Uc^n$,
\begin{align}
{1\over n}\ell_{\rm LZ}(u^n)\leq \hat{H}_k(u^n)+{b(kb+b+3)\over (1-\e_n)\log n-b}+\gamma_n,\label{eq:connections-LZ-Hk}
\end{align}
where 
\begin{align}
\e_n={2b+\log(2^b+{2^b-1\over b}\log n -2)\over \log n},
\end{align}
and $\g_n=o(1)$ does not depend  on sequence $u^n$ or $b$.

Finally, we finish this section,  by  two lemmas related to continuity properties of the entropy function and the Kullback-Leibler distance.  

\begin{lemma}[Theorem 17.3.3 in \cite{cover}]\label{lemma:bd-TV-dist}
Consider distributions $p$ and $q$ on finite alphabet $\Uc$ such that $\|p-q\|_1\leq \e$. Then,
\begin{align}
|H(p)-H(q)|\leq -\e\log \e +\e\log |\Uc|.
\end{align}
\end{lemma}

\begin{lemma}\label{lemma:dist-KL-vs-L1}
Consider distributions $p$ and $q$ over discrete set $\Uc$ such that $\|p-q\|_1\leq \e$. Further assume that $p\ll q$, and let  $q_{\min}=\min_{u\in\Uc: q(u)\neq 0} q(u)$. Then,
\begin{align}
D(p\| q)\leq -\e\log \e +\e \log |\Uc|-\e \log q_{\min}
\end{align}
\end{lemma}

\begin{IEEEproof}
Let $\Uc^*\triangleq \{u\in \Uc:\;q(u)\neq 0\}.$ Since $p\ll q$, if $q(u)=0$, then $p(u)=0$. 
Therefore, by definition
\begin{align}
D(p\| q)&= \sum_{u\in\Uc} p(u)\log {p(u)\over q(u)}\nonumber\\
&= \sum_{u\in\Uc^*} p(u)\log {p(u)\over q(u)}\nonumber\\
&= \sum_{u\in\Uc^*} p(u)\log p(u)-\sum_{u\in\Uc^*} p(u)\log q(u)\nonumber\\
&= \sum_{u\in\Uc^*} p(u)\log p(u)-\sum_{u\in\Uc^*} (p(u)-q(u)+q(u))\log q(u)\nonumber\\
&= H(q)-H(p)-\sum_{u\in\Uc^*} (p(u)-q(u))\log q(u).
\end{align}
Hence, by the triangle inequality,
\begin{align}
D(p\| q)&\leq  |H(q)-H(p)|-\sum_{u\in\Uc^*} |p(u)-q(u)|\log q(u)\nonumber\\
&\stackrel{(a)}{\leq}  -\e\log \e +\e \log |\Uc|-\sum_{u\in\Uc^*} |p(u)-q(u)|\log q(u)\nonumber\\
&\leq  -\e\log \e +\e \log |\Uc|+\log ({1\over q_{\min}})\sum_{u\in\Uc^*} |p(u)-q(u)|\nonumber\\
&\leq  -\e\log \e +\e \log |\Uc|-\e\log q_{\min}.
\end{align}
where $(a)$ and $(b)$ follow from Lemma 5 in \cite{JalaliP:14-arxiv} and $\|p-q\|_1\leq \e$, respectively. 
\end{IEEEproof}

\subsection{Useful concentration lemmas}

\begin{lemma}\label{lemma:concentration}
Consider $u^n\in \mathds{R}^n$ and $v^n\in \mathds{R}^n$ such that $\|u^n\|=\|v^n\|=1$. Let $\a\triangleq \langle u^n,v^n \rangle $. Consider matrix $A\in\mathds{R}^{m\times n}$ with i.i.d.~standard normal entries. Then, for any $\tau>0$,
\begin{align}
\P\Big({1\over m}\langle Au^n,Av^n\rangle-\langle u^n,v^n\rangle\leq -\tau\Big)\leq \ex^{m((\a-\tau)s)-{m\over 2}\ln ((1+s\a)^2-s^2)},
\end{align}
where $s>0$ is a free parameter smaller than ${1\over 1-\a}$.
\end{lemma}
\begin{IEEEproof}
Let $A_i^n$ denote the $i$-th row of matrix $A$. Then, 
\begin{align}
Au^n=\left[\begin{array}{c}
\langle A_1^n,u^n \rangle\\
\langle A_2^n,u^n \rangle\\
\vdots\\
\langle A_m^n,u^n \rangle\\
\end{array}
\right], \;\;\;\;
Av^n=\left[\begin{array}{c}
\langle A_1^n,v^n \rangle\\
\langle A_2^n,v^n \rangle\\
\vdots\\
\langle A_m^n,v^n \rangle\\
\end{array}
\right],
\end{align}
and 
\begin{align}
{1\over m}\langle Au^n,Av^n\rangle= {1\over m}\sum_{i=1}^m\langle A_i^n,u^n \rangle \langle A_i^n,v^n \rangle.
\end{align}
Let 
\begin{align}
X_i=\langle A_i^n,u^n \rangle\end{align} 
and 
\begin{align}
Y_i=\langle A_i^n,v^n \rangle.
\end{align} Since $A$ is generated by drawing its entries from an  i.i.d.~standard normal distribution, and $\|u^n\|=\|v^n\|=1$, $\{(X_i,Y_i)\}_{i=1}^m$ is a sequence of i.i.d. random vectors. To derive the joint distribution of $(X_i,Y_i)$, note that both $X_i$ and $Y_i$ are linear combination of Gaussian random variables. Therefore, they are also jointly distributed Gaussian random variables and hence it suffices to  find their first and second order moments. Note that
\begin{align}
\Ex[X_i]=\Ex[Y_i]=0,
\end{align} 
\begin{align}
\Ex[X_i^2]=\sum_{j,k}\Ex[A_{i,j}A_{i,k}]u_ju_k=\sum_{j}\Ex[A_{i,j}^2]u_j^2=\sum_{j}u_j^2=1.
\end{align}
and similarly $\Ex[Y_i^2]=1$. Also,
\begin{align}
\Ex[X_iY_i]&=\Ex[\langle A_i^n,u^n \rangle\langle A_i^n,v^n \rangle]\nonumber\\
&=\sum_{j,k}\Ex[A_{i,j}A_{i,k}]u_jv_k\nonumber\\
&=\sum_{j}\Ex[A_{i,j}^2]u_jv_j\nonumber\\
&=\langle u^n,v^n \rangle=\a.
\end{align}
Therefore, in summary, \begin{align}(X_i,Y_i)\sim \Nc\Big(
\left[
\begin{array}{c} 
0\\
0
 \end{array}\right],\left[
\begin{array}{cc} 
1&\a\\
\a& 1
 \end{array}\right]\Big).\end{align}
 For any $s'>0$, by the Chernoff bounding method, we have  
 \begin{align}
 \P\Big({1\over m}\langle Au^n,Av^n\rangle-\langle u^n,v^n\rangle\leq -\tau\Big)&=
 \P\Big({1\over m}\sum_{i=1}^m(X_iY_i-\a)\leq -\tau\Big)\nonumber\\
 &= \P\Big({s'\over m}\sum_{i=1}^m(X_iY_i-\a)\leq -s'\tau\Big)\nonumber\\
 &= \P\Big( \ex^{s'(\tau-\a)}\leq \ex^{-{s'\over m}\sum_{i=1}^m X_iY_i}\Big)\nonumber\\
 &\leq  \ex^{s'(\a-\tau)}\Ex\Big[ \ex^{-{s'\over m}\sum_{i=1}^m X_iY_i}\Big]\nonumber\\
 &= \ex^{s'(\a-\tau)}\Big(\Ex[ \ex^{-{s'\over m}X_1Y_1}]\Big)^m,\label{eq:deviation-tau}
 \end{align}
 where the last line follows because $(X_i,Y_i)$ is an i.i.d.~sequence. In the following we compute $\Ex[ \ex^{{s\over m}X_1Y_1}]$. Let $A=(X_1+Y_1)/2$ and $B=(X_1-Y_1)/2$. The, $\Ex[A]=\Ex[B]=0$ and  
 \begin{align}
 \Ex[A^2]={1+\a\over 2},
 \end{align}
 \begin{align}
 \Ex[B^2]={1-\a\over 2}
 \end{align} 
 and $\Ex[AB]=\Ex[(X_1^2-Y_1^2)/4]=0$. Therefore, $A$ and $B$ are  independent Gaussian  random variables. Therefore,
 \begin{align}
 \Ex[ \ex^{-{s'\over m}X_1Y_1}] 
 &= \Ex[ \ex^{-{s'\over m}(A+B)(A-B)}]\nonumber\\
 &= \Ex[ \ex^{-{s'\over m}A^2}] \Ex[ \ex^{{s'\over m}B^2}].
 \end{align}
 Given $Z\sim \Nc(0,\sigma^2)$, it is straightforward to show that, for $\l>-1/(2\sigma^2)$,
 \begin{align}
 \Ex[\ex^{-\l Z^2}]={1\over \sqrt{1+2\l \sigma^2}}.
 \end{align}
 Therefore, for ${s'\over m}<{1\over 1-\a}$,
  \begin{align}
 \Ex[ \ex^{{s'\over m}X_1Y_1}] &= {1\over \sqrt{(1+{s'\over m}(1+\a))(1-{s'\over m}(1-\a))}}\nonumber\\
 &= {1\over \sqrt{(1+{s'\a\over m})^2 -{s'^2\over m^2}}}.\label{eq:E-XY}
 \end{align}
 Therefore, combining \eqref{eq:deviation-tau} and \eqref{eq:E-XY}, it follows that
  \begin{align}
 \P\Big({1\over m}\langle Au^n,Av^n\rangle-\langle u^n,v^n\rangle\leq -\tau\Big)&=
 \ex^{s'(\a-\tau)}\ex^{-{m\over 2}\log\Big( (1+{s'\a\over m})^2-({s'\over m})^2 \Big)}.\label{eq:sp-s-last}
 \end{align}
 Replacing $s'/m$ with $s$ in \eqref{eq:sp-s-last} yields the desired result. 
\end{IEEEproof}

\begin{corollary}\label{cor:bound-45}
Consider $u^n\in \mathds{R}^n$ and $v^n\in \mathds{R}^n$ such that $\|u^n\|=\|v^n\|=1$.  Also, consider matrix $A\in\mathds{R}^{m\times n}$ with i.i.d.~standard normal entries. Then, 
\begin{align}
\P\Big({1\over m}\langle Au^n,Av^n\rangle-\langle u^n,v^n\rangle\leq -0.45\Big)\leq 2^{-0.05m}.
\end{align}
\end{corollary}
\begin{IEEEproof}
From Lemma \ref{lemma:concentration}, for $\a=\langle u^n,v^n\rangle$, and $s<1/(1-\a)$,
\begin{align}
\P\Big({1\over m}\langle Au^n,Av^n\rangle-\a\leq -0.45 \Big)\leq \ex^{m((\a-0.45)s)-{m\over 2}\ln ((1+s\a)^2-s^2)}=2^{-m f(\alpha,s)},
\end{align}
where
\begin{align}
f(\alpha,s)=(\log \ex)\Big({1\over 2}\ln((1+s\a)^2-s^2)-(\a-0.45)s\Big).
\end{align}
Fig.~\ref{eq:power} plots $\max_{s\in(0,{1\over 1-\a})} f(\alpha,s)$, and shows that 
\begin{align}
\min_{\alpha\in(-1,1)}\max_{s\in(0,{1\over 1-\a})} f(\alpha,s)\; \geq \; 0.05.
\end{align}

\begin{figure}[h]
\begin{center}
\includegraphics[width=10cm]{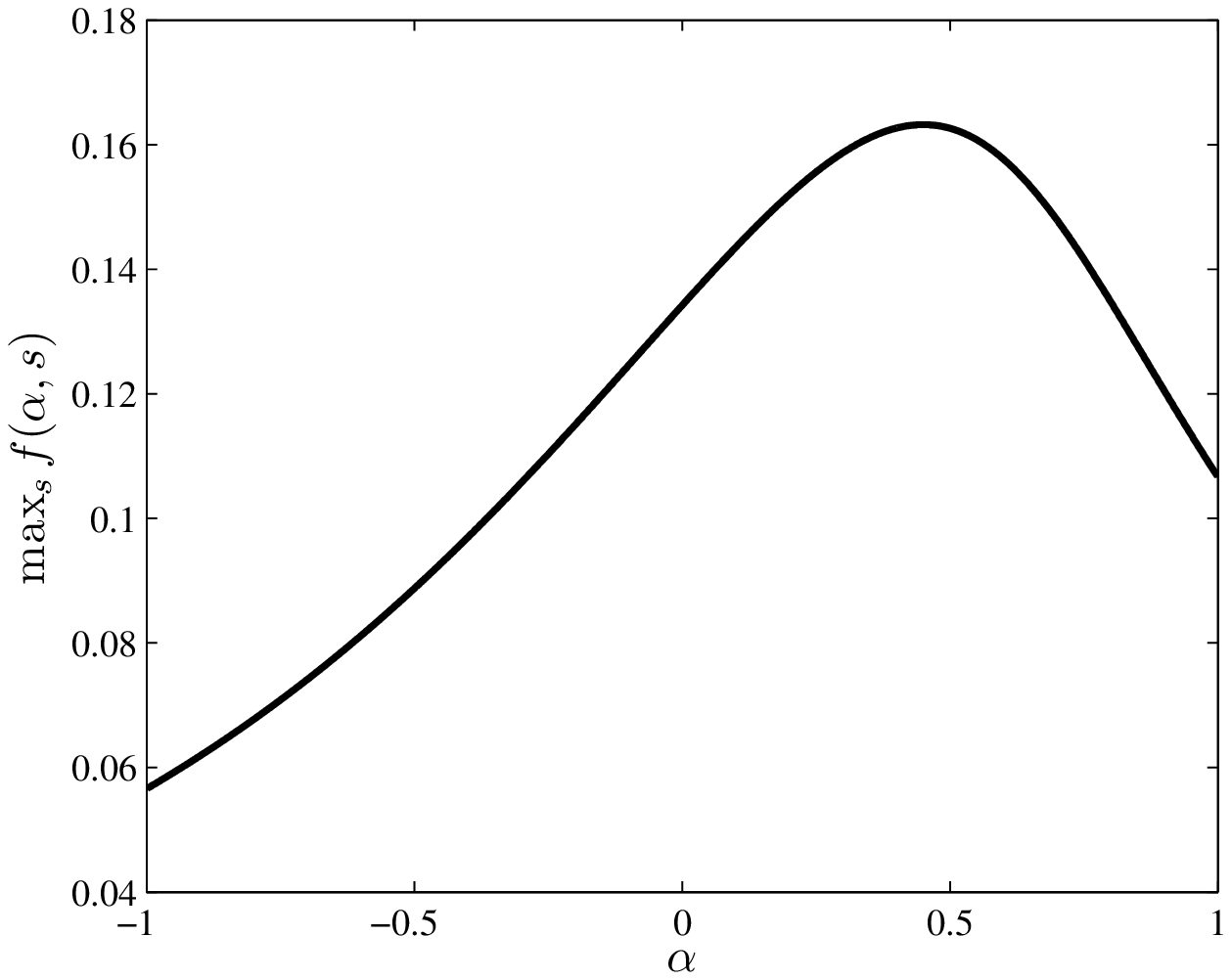}\caption{$\max_{s\in(0,{1\over 1-\alpha})} f(\a,s)$ }
\label{eq:power}
\end{center}
\end{figure}

\end{IEEEproof}

 The following two lemmas are proved in \cite{JalaliM:14-IT}.
 
\begin{lemma}[$\chi^2$ concentration]\label{lemma:chi}
Fix $\tau>0$, and let $U_i\stackrel{\rm i.i.d.}{\sim}\Nc(0,1)$, $i=1,2,\ldots,m$. Then,
\begin{align}
\P\Big( \sum_{i=1}^m  U_i^2 <m(1- \tau) \Big)  \leq {\rm e} ^{\frac{m}{2}(\tau + \ln(1- \tau))}
\end{align}
and
\begin{align}\label{eq:chisq}
\P\Big( \sum_{i=1}^m  U_i^2 > m(1+\tau) \Big)  \leq {\rm e} ^{-\frac{m}{2}(\tau - \ln(1+ \tau))}.
\end{align}
\end{lemma}

\begin{lemma}\label{lemma:gaussian-vectors}
Consider $U^n$ and $V^n$, where, for each $i$,  $U_i$ and $V_i$ are two independent  standard normal random variables.  Then the distribution of  $\langle U^n,V^n \rangle=\sum_{i=1}^nU_iV_i$ is the same as the distribution of $\|U^n\|G$, where $G\sim\Nc(0,1)$ is independent of $\|U^n\|_2$.
\end{lemma}



\subsection{Proof of Lemma \ref{lemma:Psi-dist}} \label{proof:lemma2}

Before presenting the proof, we establish some preliminary results. 
Consider an analog process $\Xbbf=\{X_i\}$ with alphabet $\Xc=[l,u]$, where $l,u\in\mathds{R}$. Let process $\Zbbf=\{Z_i\}$ denote the $b$-bit quantized version of process $\Xbbf$. That is, $Z_i=[X_i]_b$, and the alphabet of process $\Zbbf$ is $\Zc=\Xc_b=\{[x]_b: x\in\Xc_b\}$. 
For $i=1,\ldots,k+g$, define a sequence of length $t$, $\{S^{(i)}_j\}_{j=1}^t$ over super-alphabet $\Zc^k$ as follows. For $i=1$,   $\{S^{(1)}_j\}_{j=1}^t$ is defined as 
\begin{align}
&\underbrace{Z_1,\ldots,Z_k}_{S_1^{(1)}},Z_{k+1},\ldots,Z_{k+g},\underbrace{Z_{k+g+1},\ldots,Z_{2k+g+1}}_{S_2^{(1)}},\\
&{Z_{2k+g+2},\ldots \ldots}, \underbrace{Z_{(t-1)(k+g)+1},\ldots,Z_{(t-1)(k+g)+k}}_{S_t^{(1)}},Z_{t(k+g)-g+1},\ldots,Z_n.
\end{align}
Similarly, $\{S^{(i)}_j\}_{j=1}^t$, $i=1,\ldots,k+g$, is defined by starting the grouping of the symbols at $Z_i$. In other words, 
\begin{align}
S^{(i)}_j\triangleq Z_{(k+g)(j-1)+i}^{(k+g)(j-1)+i+k-1}.
\end{align} 
For instance, the sequence $\{S^{(k+g)}_j\}_{j=1}^t$, which  corresponds to the largest shift at beginning, is defined as 
\begin{align}
&Z_1,\ldots,Z_{k+g-1},\underbrace{Z_{k+g},\ldots,Z_{2k+g-1}}_{S_1^{(k)}},Z_{2k+g},\ldots,Z_{2k+2g-1},\\
&\underbrace{Z_{2k+2g},\ldots,Z_{3k+2g-1}}_{S_2^{(k)}},Z_{3k+2g}\ldots.
\end{align}
This definition  implies that $n$  satisfies
\begin{align}
(t-1)(k+g)+2k+g-1\leq n<(t-1)(k+g)+2k+g-1+k+g.\label{eq:bound-n}
\end{align}
That is, $t$ is the only integer in the $({n-2k-g+1\over k+g},{n-k+1\over k+g} ]$
interval, or in other words, $t=\lfloor{n-k+1\over k+g}\rfloor$.

Before we prove Lemma \ref{lemma:Psi-dist}, we prove the following auxiliary lemma. 

\begin{lemma}\label{lemma:overlapping-non-overlapping}
For any given $\e>0$ and any positive integers $g$ and $k$ such that $4(k+g)/(n-k)<\e$, if $\|\hat{p}^{(k)}(\cdot|Z^n)-\mu_k^{(b)}\|_1\geq \e$, then that there exists $i=1,\ldots,k+g$, such that
\begin{align}
\|\hat{p}^{(1)}(\cdot|S^{(i),t})-\mu_k^{(b)}(\cdot)\|_1 \geq {\e\over 2},
\end{align}
where  $S^{(i),t}$ denotes the sequence $S^{(i)}_1, S^{(i)}_2, \ldots, S^{(i)}_t$. 
\end{lemma}

Note that in Lemma \ref{lemma:overlapping-non-overlapping}, $\hat{p}^{(1)}(\cdot|S^{(i),t})$ denotes the standard first order empirical distribution of the sup-alphabet sequence  $S^{(i),t}$, \ie for $a^k\in\Xc^k$, 
\begin{align}
\hat{p}^{(1)}(a^k|S^{(i),t})={|\{j: S^{(i)}_j=a^k, 1\leq j \leq t\}|\over t}.
\end{align}

\begin{IEEEproof}
Note that by definition, for any $a^k\in\Zc^k$,
\begin{align}
\hat{p}^{(k)}(a^k|Z^n)&={1\over n-k}\sum_{i=k}^n \ind_{Z_{i-k+1}^i=a^k}\nonumber\\
&={1\over n-k}\sum_{i=k}^n \ind_{Z_{i-k+1}^i=a^k}\nonumber\\
&={1\over n-k}\Big(\sum_{i=1}^{k+g}\sum_{j=1}^t \ind_{S^{(i)}_j=a^k}+\sum_{t(k+g)+k}^n \ind_{Z_{i-k+1}^i=a^k}\Big)\nonumber\\
&={t\over n-k}\sum_{i=1}^{k+g}\Big({1\over t}\sum_{j=1}^t \ind_{S^{(i)}_j=a^k}\Big)+{1\over n-k}\sum_{t(k+g)+k}^n \ind_{Z_{i-k+1}^i=a^k}\nonumber\\
&={t\over n-k}\sum_{i=1}^{k+g}\hat{p}^{(1)}(a^k|S^{(i),t})+{1\over n-k}\sum_{t(k+g)+k}^n \ind_{Z_{i-k+1}^i=a^k}.\label{eq:emp-total-to-overlapping}
\end{align}
Therefore,
\begin{align}
&\|\hat{p}^{(k)}(\cdot|Z^n)-\mu_k^{(b)}\|_1=\sum_{a^k}|\hat{p}^{(k)}(a^k|Z^n)-\mu_k^{(b)}(a^k)|\nonumber\\
&\stackrel{(a)}{=}\sum_{a^k}\Big|{t\over n-k}\sum_{i=1}^{k+g}\hat{p}^{(1)}(a^k|S^{(i),t})+{1\over n-k}\sum_{t(k+g)+k}^n \ind_{Z_{i-k+1}^i=a^k}-\mu_k^{(b)}(a^k)|\nonumber\\
&\stackrel{(b)}{=}\sum_{a^k}\Big|{t\over n-k}\sum_{i=1}^{k+g}\hat{p}^{(1)}(a^k|S^{(i),t})-\mu_k^{(b)}(a^k)\Big|+{1\over n-k}\sum_{a^k}\sum_{t(k+g)+k}^n \ind_{Z_{i-k+1}^i=a^k}\nonumber\\
&\stackrel{(c)}{=}\sum_{a^k}\Big|{t\over n-k}\sum_{i=1}^{k+g}\hat{p}^{(1)}(a^k|S^{(i),t})-\mu_k^{(b)}(a^k)\Big|+{n-t(k+g)-k+1\over n-k},\label{eq:ell1-phat-k-mu-k-vs-phat-shifted-extra}
\end{align}
where $(a)$ follows from \eqref{eq:emp-total-to-overlapping}, $(b)$ follows from the triangle inequality and $(c)$ holds because 
\begin{align}
\sum_{a^k} \ind_{Z_{i-k+1}^i=a^k}=1.
\end{align}
But since $n$ satisfies the bounds of \eqref{eq:bound-n}, the last term on the right hand side of \eqref{eq:emp-total-to-overlapping} can be upper-bounded as
\begin{align}
{n-t(k+g)-k+1\over n-k}\leq {k+g\over n-k}.
\end{align}
Therefore, from \eqref{eq:ell1-phat-k-mu-k-vs-phat-shifted-extra} we have
\begin{align}
\|\hat{p}^{(k)}(\cdot|Z^n)-\mu_k^{(b)}\|_1& \leq \sum_{a^k}\Big|{t\over n-k}\sum_{i=1}^{k+g}\hat{p}^{(1)}(a^k|S^{(i),t})-\mu_k^{(b)}(a^k)\Big|+ {k+g\over n-k}.\label{eq:dist-phat-k-avg-phat-1}
\end{align}

On the other hand, again by the triangle inequality, 
\begin{align}
&\sum_{a^k}\Big|{t\over n-k}\sum_{i=1}^{k+g}\hat{p}^{(1)}(a^k|S^{(i),t})-\mu_k^{(b)}(a^k)\Big|\nonumber\\
&=\sum_{a^k}\Big|{t\over n-k}\sum_{i=1}^{k+g}\left(\hat{p}^{(1)}(a^k|S^{(i),t})-\mu_k^{(b)}(a^k)\right)+({t(k+g)\over n-k}-1)\mu_k^{(b)}(a^k)\Big| \nonumber\\
&\leq \sum_{a^k}\Big|{t\over n-k}\sum_{i=1}^{k+g}\left(\hat{p}^{(1)}(a^k|S^{(i),t})-\mu_k^{(b)}(a^k)\right)\Big|+|{t(k+g)\over n-k}-1|\sum_{a^k} \mu_k^{(b)}(a^k) \nonumber\\
&= \sum_{a^k}\Big|{t\over n-k}\sum_{i=1}^{k+g}\left(\hat{p}^{(1)}(a^k|S^{(i),t})-\mu_k^{(b)}(a^k)\right)\Big|+|{t(k+g)\over n-k}-1|\nonumber\\
&\leq {t\over n-k} \sum_{i=1}^{k+g} \sum_{a^k}\Big|\hat{p}^{(1)}(a^k|S^{(i),t})-\mu_k^{(b)}(a^k)\Big|+|{t(k+g)\over n-k}-1|\nonumber\\
&= {t\over n-k} \sum_{i=1}^{k+g} \|\hat{p}^{(1)}(\cdot|S^{(i),t})-\mu_k^{(b)}(\cdot)\|_1+|{t(k+g)\over n-k}-1|.\label{eq:dist-phat-k-avg-phat-2}
\end{align}
Therefore, combining \eqref{eq:ell1-phat-k-mu-k-vs-phat-shifted-extra} and \eqref{eq:dist-phat-k-avg-phat-2} yields 
\begin{align}
\|\hat{p}^{(k)}(\cdot|Z^n)-\mu_k^{(b)}\|_1& \leq  {t\over n-k} \sum_{i=1}^{k+g} \|\hat{p}^{(1)}(\cdot|S^{(i),t})-\mu_k^{(b)}(\cdot)\|_1+|{t(k+g)\over n-k}-1|+{k+g\over n-k}.\label{eq:dist-phat-k-avg-phat-3}
\end{align}
Since by construction $t(k+g)\leq n-k$, we have $t/(n-k)\leq 1/(k+g)$. Therefore, it follows  from \eqref{eq:dist-phat-k-avg-phat-3} that:
\begin{align}
\|\hat{p}^{(k)}(\cdot|Z^n)-\mu_k^{(b)}\|_1& \leq  {1\over k+g} \sum_{i=1}^{k+g} \|\hat{p}^{(1)}(\cdot|S^{(i),t})-\mu_k^{(b)}(\cdot)\|_1+1-{t(k+g)\over n-k}+{k+g\over n-k}\nonumber\\
&=   {1\over k+g} \sum_{i=1}^{k+g} \|\hat{p}^{(1)}(\cdot|S^{(i),t})-\mu_k^{(b)}(\cdot)\|_1+{n-t(k+g)+g\over n-k}.\label{eq:dist-phat-k-avg-phat-4}
\end{align}
Notice that if 
\begin{align}
{n-t(k+g)+g\over n-k}\leq {\e\over 2},\label{eq:cond-k-g-epsilon}
\end{align}
and 
\begin{align}
\|\hat{p}^{(1)}(\cdot|S^{(i),t})-\mu_k^{(b)}(\cdot)\|_1\leq {\e\over 2},
\end{align}
for all $i$, then, from \eqref{eq:dist-phat-k-avg-phat-4}, $\|\hat{p}^{(k)}(\cdot|Z^n)-\mu_k^{(b)}\|_1\leq {\e}$. But, if $4(k+g)/(n-k)<\e$, since  since $t=\lfloor {n-k+1\over k+g} \rfloor$, then
\eqref{eq:cond-k-g-epsilon} holds. This means that,  to have   $\|\hat{p}^{(k)}(\cdot|Z^n)-\mu_k^{(b)}\|_1> {\e}$, we need $\|\hat{p}^{(1)}(\cdot|S^{(i),t})-\mu_k^{(b)}(\cdot)\|_1> {\e\over 2}$, for  at least one $i$ in $\{1,\ldots,k+g\}$.

\end{IEEEproof}

Now we can discuss the proof of Lemma \ref{lemma:Psi-dist}. The proof is a  straightforward extension of Lemma III.1.3  in \cite{Shields}. However, we include a summary of the proof for completeness.
By Lemma \ref{lemma:overlapping-non-overlapping}, if $4(k+g)/(n-k)<\e$, then $\|\hat{p}^{(k)}(\cdot|Z^n)-\mu_k^{(b)}\|_1\geq \e$ implies that there exists $i\in\{1,\ldots,k+g\}$ such that
\begin{align}
\|\hat{p}^{(1)}(\cdot|S^{(i),t})-\mu_k^{(b)}(\cdot)\|_1\geq {\e\over 2}.
\end{align}
We next bound the probability that the above event happens. 
For each $i\in\{1,\ldots,k+g\}$, define event $\Ec^{(i)}$ as follows
\begin{align}
\Ec^{(i)}\triangleq \{D_{\rm KL}(\hat{p}^{(1)}(\cdot|S^{(i),t}),\mu_k^{(b)})> \e^2/2\}.
\end{align}
By the Pinsker's inequality, for any $i$, 
\begin{align}
\|\hat{p}^{(1)}(\cdot|S^{(i),t})-\mu_k^{(b)}\|_1 \leq \sqrt{(2\ln 2) D_{\rm KL}(\hat{p}^{(1)}(\cdot|S^{(i),t}),\mu_k^{(b)}(\cdot))}.
\end{align} Therefore, if $D_{\rm KL}(\hat{p}^{(1)}(\cdot|S^{(i),t}),\mu_k^{(b)}(\cdot))\leq c\e^2/4$, where as defined earlier $c=1/(2\ln 2)$, then $\|\hat{p}^{(1)}(\cdot|S^{(i),t})-\mu_k^{(b)}(\cdot)\|_1 \leq \e/2$. This implies that
\begin{align}
\P(\|\hat{p}^{(1)}(\cdot|S^{(i),t})-\mu_k^{(b)}\|_1 >{\e\over 2} )\leq \P\Big(D_{\rm KL}(\hat{p}^{(1)}(\cdot|S^{(i),t}),\mu_k^{(b)}) >{c\e^2\over 4}\Big).
\end{align}
On the other hand, for $S^{(i),t}=s^t$, where $s^t\in (\Zc^k)^t$, we have
\begin{align}
\P(S^{(i),t}=s^t)&=\P\Big(Z_{(k+g)(j-1)+i}^{(k+g)(j-1)+i+k-1}=s_j, \;j=1,\ldots,t\Big)\nonumber\\
&\stackrel{(a)}{\leq} \Psi^t(b,g) \prod_{j=1}^t \P\Big(Z_{(k+g)(j-1)+i}^{(k+g)(j-1)+i+k-1}=s_j\Big),
\end{align}
where $(a)$ follows from applying condition \eqref{eq:cond-Psi-b-g} $t$ times. But, by the standard method of types techniques \cite{csiszar2011information},  we have
\begin{align}
 \prod_{j=1}^t\P\Big(Z_{(k+g)(j-1)+i}^{(k+g)(j-1)+i+k-1}=s_j\Big)&= 2^{-t(\hat{H}_1(s^t)+D_{\rm KL}(\hat{p}^{(1)}(\cdot|s^t),\mu_k^{(b)}))}.
\end{align}
Therefore, if $D_{\rm KL}(\hat{p}^{(1)}(\cdot|s^t),\mu_k^{(b)}) >{c\e^2\over 4}$, then
\begin{align}
 \prod_{j=1}^t\P\Big(Z_{(k+g)(j-1)+i}^{(k+g)(j-1)+i+k-1}=s_j\Big)&\leq 2^{-t\hat{H}_1(s^t)-c\e^2t/4}.
\end{align}
Hence, 
\begin{align}
\P&\Big(D_{\rm KL}(\hat{p}^{(1)}(\cdot|S^{(i),t}),\mu_k^{(b)})>{c\e^2\over 4}\Big)\nonumber\\
&=\sum_{s^t:D_{\rm KL}(\hat{p}^{(1)}(\cdot|s^t),\mu_k^{(b)}) >{c\e^2\over 4} }\P(S^{(i),t}=s^t)\nonumber
\\
&\leq 2^{-c\e^2t/4}\Psi^t(b,g) \sum_{s^t:D_{\rm KL}(\hat{p}^{(1)}(\cdot|s^t),\mu_k^{(b)})  >{c\e^2\over 4} }2^{-t\hat{H}_1(s^t)}\nonumber\\
&\leq 2^{-c\e^2t/4} \Psi^t(b,g) \sum_{s^t }2^{-t\hat{H}_1(s^t)}.
\end{align}
Since $\sum_{s^t }2^{-t\hat{H}_1(s^t)}$ can be proven to be smaller than the total number of types of sequences $s^t\in(\Zc^k)^t$, we have $\sum_{s^t }2^{-t\hat{H}_1(s^t)}\leq (t+1)^{|\Zc|^k}$. This upper bound combined by the union bound on $\Ec^{(i)}$, $i=1,\ldots,k+g$, yields the desired result.  


\subsection{Proof of Lemma \ref{lemma:Psi-Markov}} \label{proof:lemma3}
For any $u^{\ell_1}\in\Zc^{\ell_1}$, $v^{g}\in\Zc^{g}$, and $w^{\ell_2}\in\Zc^{\ell_2}$, we have
\begin{align}\label{eq:quant-markov}
&\P(Z^{\ell_1+g+\ell_2}=[u^{\ell_1}v^{g}w^{\ell_2}])\nonumber\\
&\leq \sum_{v^{g}\in\Zc^g} \P(Z^{\ell_1+g+\ell_2}=[u^{\ell_1}v^{g}w^{\ell_2}])\nonumber\\
&= \P(Z^{\ell_1}=u^{\ell_1}, Z_{\ell_1+g+1}^{\ell_1+g+\ell_2}=w^{\ell_2})\nonumber\\
  &= \P(Z^{\ell_1}=u^{\ell_1}) \P(Z_{\ell_1+g+1}=w_1 |  Z^{\ell_1}= u^{\ell_1}) \nonumber\\
  &\;\;\;\times \P(Z_{\ell_1+g+2}^{\ell_1+g+\ell_2}=w_2^{\ell_2}|Z_{\ell_1+g+1}=w_1,Z^{\ell_1}=u^{\ell_1})\nonumber\\
  &= \P(Z^{\ell_1}=u^{\ell_1})\int_{\Xc} \P(Z_{\ell_1+g+1}=w_1 | X_{\ell_1}= x_{\ell_1}, Z^{\ell_1}= u^{\ell_1})  d\mu(x_{\ell_1}| Z^{\ell_1}= u^{\ell_1})  \nonumber\\
  &\;\;\;\; \times \P(Z_{\ell_1+g+2}^{\ell_1+g+\ell_2}=w_2^{\ell_2}|Z_{\ell_1+g+1}=w_1,Z^{\ell_1}=u^{\ell_1})\nonumber\\
    & \stackrel{(a)}{=} \P(Z^{\ell_1}=u^{\ell_1})\int_{\Xc} \P(Z_{\ell_1+g+1}=w_1 | X_{\ell_1}= x_{\ell_1})  d\mu(x_{\ell_1}| Z^{\ell_1}= u^{\ell_1})  \nonumber\\
  &\;\;\;\; \times \P(Z_{\ell_1+g+2}^{\ell_1+g+\ell_2}=w_{2}^{\ell_2}|Z_{\ell_1+g+1}=w_1,Z^{\ell_1}=u^{\ell_1})\nonumber\\
    & \stackrel{(b)}{=} \mu_b(u^{\ell_1})\Big({\int_{\Xc}{K^{g+1}(x_{\ell_1},w_1)}  d\mu(x_{\ell_1}| u^{\ell_1}) }\Big) \mu_b(w_2^{\ell_2}|w_1,u^{\ell_1})\nonumber\\
   & = \mu_b(u^{\ell_1})\Big({\int_{\Xc^l}{K^{g+1}(x_{\ell_1},w_1)\over \pi(w_1)}  d\mu(x_{\ell_1}| u^{\ell_1}) }\Big) \pi(w_1)\mu_b(w_2^{\ell_2}|w_1,u^{\ell_1})\nonumber\\
  &\leq  \mu_b(u^{\ell_1}) \pi(w_1)\mu_b(w_2^{\ell_2}|w_1,u^{\ell_1})\Big(\sup_{x: [x]_b=u_{\ell_1}} {{K^{g+1}(x,w_1)}  \over \pi(w_1)}\Big)\nonumber\\
   &\leq  \mu_b(u^{\ell_1}) \pi(w_1)\mu_b(w_2^{\ell_2}|w_1,u^{\ell_1})\Big(\sup_{(x,z)\in\Xc\times \Zc} {{K^{g+1}(x,z)}  \over \pi(z)}\Big)\nonumber\\
      &=  \mu_b(u^{\ell_1}) \pi(w_1)\mu_b(w_2^{\ell_2}|w_1,u^{\ell_1})\Psi_1(b,g),
\end{align}
 where (a) holds because $\Xbbf$ is a first order Markov chain  and  in (b), 
 \begin{align}
 \mu_b(u^{\ell_1})= \P(Z^{\ell_1}=u^{\ell_1}),
 \end{align} 
 \begin{align}
 \mu_b(w_2^{\ell_2}|w_1,u^{\ell_1})=\P(Z_{\ell_1+g+2}^{\ell_1+g+\ell_2}=w_2^{\ell_2}|Z_{\ell_1+g+1}=w_1,Z^{\ell_1}=u^{\ell_1}),
 \end{align} 
 and $\mu(x_{\ell_1}| u^{\ell_1})$ denotes the  probability measure of $X_{\ell_1}$ conditioned on $Z^{\ell_1}=u^{\ell_1}$. Also, since the Markov chain is a stationary process, we have  
 \begin{align}
 K^{g+1}(x_{\ell_1},w_1)=\P([X_{g+1}]_b=w_1|X_0=x_{\ell_1}).
 \end{align}

Another term in  \eqref{eq:quant-markov} is $ \pi(w_1)\mu_b(w_{2}^{\ell_2}|w_1,u^{\ell_1})$. Since $\pi(w^{\ell_2})= \pi(w_1) \pi(w_{2}^{\ell_2}|w_1)$, we have
\begin{align}
 \pi(w^l)\mu_b(w_{l+1}^{\ell_2}|w^l,u^{\ell_1})&= \pi(w^{\ell_2}){\mu_b(w_{l+1}^{\ell_2}|w^l,u^{\ell_1})\over \pi(w_{l+1}^{\ell_2}|w^l) }.\label{eq:bayes-rule}
\end{align}
But 
\begin{align}
\mu_b(w_2^{\ell_2}|w_1,u^{\ell_1})&=\int \mu_b(w_{2}^{\ell_2}|x,w_1,u^{\ell_1})d\mu(x|w_1,u^{\ell_1})\nonumber\\
&=\int \mu_b(w_2^{\ell_2}|x)d\mu(x|w_1,u^{\ell_1})\label{eq:int-Markov}
\end{align}
where the second equality holds because $(Z^{\ell_1},Z_{\ell_1+1})\to X_{\ell_1+1}\to Z_{\ell_1+2}^{\ell_1+l+\ell_2}$ forms a Markov chain. Therefore,
\begin{align}
{\mu_b(w_{2}^{\ell_2}|w_1,u^{\ell_1})\over \pi(w_{2}^{\ell_2}|w_1) }
&={\int \mu_b(w_{2}^{\ell_2}|x)d\mu(x|w_1,u^{\ell_1})\over \pi(w_2^{\ell_2}|w_1) },\nonumber\\
&\stackrel{(a)}{\leq } {\Big(\sup\limits_{[x]_b=w_1}\mu_b(w_2^{\ell_2}|x)\Big)\int d\mu(x|w_1,u^{\ell_1})\over \pi(w_{2}^{\ell_2}|w_1) },\nonumber\\
&\stackrel{(b)}{= } \sup\limits_{[x]_b=w_1}{\pi(w_{2}^{\ell_2}|x)\over \pi(w_{2}^{\ell_2}|w_1)}\nonumber\\
&\leq \sup_{(x,w^{\ell_2}): [x]_b=w_1} {\pi(w_{2}^{\ell_2}|x)\over \pi(w_{2}^{\ell_2}|w_1)}\nonumber\\
&\leq \Psi_2(b),\label{eq:bd-realted-psi2}
\end{align}
where (a) and (b) hold  because  $\mu(x|w_1,u^{\ell_1})$ is only non-zero when $x$ is such that $[x]_b=w_1$, and $\int d\mu(x|w_1,u^{\ell_1})=1$, respectively. 
Finally, combining \eqref{eq:quant-markov}, \eqref{eq:bayes-rule} and \eqref{eq:bd-realted-psi2}  yields the desired result.

We next prove that, for a fixed $b$, $\Psi_1(b,g)$ is non-increasing function of $g$. For any $x\in\Xc$ and $z\in\Xc_b$, we have 
\begin{align}
{K^{g+1}(x,z) \over \pi(z)}   &={\P([X_{g+1}]_b=z|X_0=x) \over \P([X_{g+1}]_b=z)}  \nonumber\\
&={\int \P([X_{g+1}]_b=z|X_1=x',X_0=x)d\mu(x'|X_0=x) \over \P([X_{g+1}]_b=z)}  \nonumber\\
&\stackrel{(a)}{=}{\int \P([X_{g}]_b=z|X_0=x')d\mu(x'|X_0=x) \over \P([X_{g}]_b=z)}  \nonumber\\
&=\sup_{x'\in\Xc} {\P([X_{g}]_b=z|X_0=x') \over \P([X_{g}]_b=z)} \int d\mu(x'|X_0=x) \nonumber\\
&\stackrel{(b)}{\leq} \Psi_1(g,b),\label{eq:Kg-plus-1}
\end{align}
where $(a)$ follows because of the Markovity and stationarity  assumptions and $(b)$ follows because $\int d\mu(x'|X_0=x)=1$. 
Since the right hand side of \eqref{eq:Kg-plus-1} only depends on $g$ and $b$, taking the supremum of the left hand side over $(x,z)\in\Xc\times\Xc_b$ proves that 
\begin{align}
\Psi_1(g+1)\leq \Psi_1(g).
\end{align}
Furthermore, since $\Xbbf$ is assumed to be an aperiodic Markov chain, \\  $\lim_{g\to\infty} K^{g}(x,z)=\pi(z)$, for all $x$ and $z$. Therefore, $\Psi_1(g)$ converges to one, as $g\to\infty$.



\subsection{Proof of Theorem \ref{thm:exp-rate-markov}}\label{proof:thm6}
Define $\Psi(b,g)=\Psi_1(b,g)\Psi_2(b)$. Then it follows from  Lemmas \ref{lemma:Psi-dist}  and  \ref{lemma:Psi-Markov} that, given $\e>0$, for  any positive integers $g$ and $k$ that satisfy $4(k+g)/(n-k)<\e$,
\begin{align}
\P(\|\hat{p}^{(k)}(\cdot|Z^n)-\mu_k^{(b)}\|_1\geq \e)\leq (k+g)\Psi_1^t(b,g)\Psi_2^t(b)(t+1)^{|\Zc|^k}2^{-c\e^2t/4}.\label{eq:ell-1-main}
\end{align}
where $t=\lfloor{n-k+1\over k+g}\rfloor$ and $c=1/(2\ln 2)$.
Since by assumption  $\lim_{b\to\infty}\Psi_2(b)=1$, there exists $b_{\e}$ such that for all $b\geq b_{\e}$, $\Psi_2(b)\leq 2^{c\epsilon^2/16}$. 
But $ b_n=\lceil r\log\log n \rceil$ is a diverging sequence of $n$. Therefore, there exists $n_{\e}>0$, such that for all $n\geq n_{\e}$,  
\begin{align}
\Psi_2(b_n)\leq 2^{c\epsilon^2/16}.\label{eq:bd-Psi-2}
\end{align}
On the other hand, by the theorem's assumption, there exists a sequence $g=g_n$, where $g=o(n)$, such that $\lim_{n\to\infty}\Psi_1(b_n,g_n)=1 $. Therefore, there exists $n'_{\e}$ such that for all $n\geq n'_{\e}$,
\begin{align}
\Psi_1(b_n,g_n)\leq 2^{c\epsilon^2/16}.\label{eq:bd-Psi-1}
\end{align}
Moreover, since $g=g_n=o(n)$ and $k$ is fixed, there exists $n''_{\e}>0$ such that for all $n\geq n''_{\e}$,
\begin{align}
{4(k+g_n) \over n-k}<\e.\label{eq:k-g-n}
\end{align}
Therefore, for $n>\max(n_{\e},n'_{\e},n''_{\e})$, from \eqref{eq:ell-1-main}, \eqref{eq:bd-Psi-2}, \eqref{eq:bd-Psi-1} and \eqref{eq:k-g-n}, we have
\begin{align}
\P(\|\hat{p}^{(k)}(\cdot|Z^n)-\mu_k^{(b)}\|_1\geq \e)&\leq (k+g)\Psi_1^t(b,g)\Psi_2^t(b)(t+1)^{|\Zc|^k}2^{-c\e^2t/4}\nonumber\\
&\leq  (k+g)(t+1)^{|\Zc|^k}2^{-tc\e^2/8}\nonumber\\
&\leq  (k+g)n^{|\Zc|^k}2^{-tc\e^2/8},
\end{align}
where the last line follows from the fact that $t+1\leq n$. But  since $t=\lfloor{n-k+1\over k+g}\rfloor$, $t\geq {n-k\over k+g}-1$. Hence,
\begin{align}
\P(\|\hat{p}^{(k)}(\cdot|Z^n)-\mu_k^{(b)}\|_1\geq \e)&\leq 2^{c\e^2(1+k/(k+g))/8} (k+g)n^{|\Zc|^k}2^{-{c \e^2n\over 8(k+g)}}\nonumber\\
&\leq 2^{c\e^2/4} (k+g)n^{|\Zc|^k}2^{-{c \e^2n\over 8(k+g)}},
\end{align}
which is the desired result.

\subsection{Proof of Theorem \ref{thm:Markov-Psi1-Psi2}}\label{proof:thm7}

For each $i$, let random variable $J_i$ be an indicator of a jump at time $i$. That is,
\begin{align}
J_i=\ind_{X_i\neq X_{i-1}}.
\end{align}
Consider $x\in \Xc$ and $z\in\Xc_b$. Then, by definition,
\begin{align}
{K^g(x,z)\over \pi(z)}&={\P([X_g]_b=z|X_0=x)\over \P([X_0]_b=z)}.
\end{align}
But,  
\begin{align}
\P([X_g]_b=z|X_0=x)&= \sum_{d^g\in\{0,1\}^g}\P([X_g]_b=z,J^g=d^g|X_0=x)\nonumber\\
&\stackrel{(a)}{=}\sum_{d^g\in\{0,1\}^g}\P([X_g]_b=z|J^g=d^g,X_0=x)\P(J^g=d^g).
\end{align}
where $(a)$ follows from the independence of the jump events and the value of the Markov process at each time. Now if there is a jump between time $1$ and time $g$, then by definition of the transition probabilities the value of  $[X_g]_b$ become independent of $[X_1]_b$ and also the jumps pattern. In other words, for any $J^g\neq (0,\ldots,0)$,
\begin{align}
\P([X_g]_b=z|J^g=d^g,X_0=x)=\P([X_g]_b=z)=\P([X_0]_b=z),
\end{align}
where the last equality follows from the stationarity  of the Markov process. But $J^g\neq (0,\ldots,0)$ means that there has been no jump from time $0$ upto time $g$, and therefore $X_g=X_0$. This implies that 
\begin{align}
\P([X_g]_b=z|J^g=0^g,X_0=x)=\ind_{z=[x]_b}.
\end{align}
Since $J^g= (0,\ldots,0)=(1-p)^g$, combining the intermediate steps, its follows that  
\begin{align}
\P([X_g]_b=z|X_0=x)&= (1-(1-p)^g)\P([X_0]_b=z)+(1-p)^g\ind_{z=[x]_b},
\end{align}
and as a result 
\begin{align}
{K^g(x,z)\over \pi(z)}&=(1-(1-p)^g)+ {(1-p)^g\ind_{z=[x]_b}\over \P([X_0]_b=z)}.
\end{align}
But given that by our assumption $f(x)\geq f_{\min}$,  $\P([X_0]_b=z)\geq f_{\min} 2^{-b}$. Therefore,
\begin{align}
\Psi_1(b,g)& =\sup_{(x,z)\in\Xc\times \Xc_b} {K^g(x,z)\over \pi(z)}\nonumber\\
&\leq (1-(1-p)^g)+ {(1-p)^g 2^b\over f_{\min}}.\label{eq:bd-Psi-1-Markov}
\end{align}
Ffor  $b=b_n=\lceil r\log\log n \rceil $ and $g=g_n=\lfloor \g r\log\log n \rfloor$,  we have
\begin{align}
\log ((1-p)^g 2^b)=g\log (1-p)+b\\
\leq  r(\gamma \log(1-p) +1)\log\log n. 
\end{align}
But  since  $\g>-{1\over \log(1-p)}$, $\gamma \log(1-p) +1<0$, which from \eqref{eq:bd-Psi-1-Markov} proves the desired result, \ie  $\lim_{n\to\infty}\Psi_1(b_n,g_n)=1$.

It is easy to check that due to its special distribution, the quantized version of process $\Xbbf$ is also a first-order Markov process. Therefore, from \eqref{eq:Psi2-def}, we have
\begin{align}
\Psi_2(b)&=\sup_{(x,w^2)\in\Xc\times \Zc^{2}: [x]_b=w_1} {\pi(w_{2}|x)\over \pi(w_{2}|w_1)}\nonumber\\
&=\sup_{(x,w_2)\in\Xc\times \Zc} {\P([X_2]_b=w_2|X_1=x)\over \P([X_2]_b=w_2|[X_1]_b=[x]_b)}.
\end{align}
But
\begin{align}
\P([X_2]_b=w_2|X_1=x)&=\P([X_2]_b=w_2,J_2=1|X_1=x)p\nonumber\\
&\;\;\;+\P([X_2]_b=w_2,J_2=0|X_1=x)(1-p)\nonumber\\
&=p\P([X_2]_b=w_2)+(1-p)\ind_{w_2=[x]_b}.
\end{align}
and similarly 
\begin{align}
\P([X_2]_b=w_2|[X_1]_b=[x]_b)&=\P([X_2]_b=w_2,J_2=1|[X_1]_b=[x]_b)p\nonumber\\
&\;\;\;+\P([X_2]_b=w_2,J_2=0|[X_1]_b=[x]_b)(1-p)\nonumber\\
&=p\P([X_2]_b=w_2)+(1-p)\ind_{w_2=[x]_b},
\end{align}
which proves that $\Psi_2(b)=1$, for all $b$.


\subsection{Proof of Theorem \ref{thm:mainresult-Q-MAP}}\label{sec:proof-thm2}

By  definition 
\begin{align}
\bar{d}_k(\Xbbf)=\limsup_{b\to\infty} {H([X_{k+1}]_b|[X^k]_b)\over b},
\end{align}
and therefore, for any $\delta_1>0$, there exists $b_{\d_1}$ such that for all $b\geq b_{\d_1}$,  
${H(X_{k+1}]_b|[X^k]_b)\over  b} \leq \bar{d}_k(\Xbbf) + \d_1.
$
Since $b=b_n=\lceil r\log\log n \rceil$ converges to infinity as $n\to\infty$, for all $n$ large enough, $b=b_n>b_{\d_1}$, and as a result 
\begin{align}
 {H([X_{k+1}]_b|[X^k]_b)\over b} \leq \bar{d}_k(\Xbbf) + \d_1.
\end{align}
For the rest of the proof, assume that $n$ is larges enough such that $b_n>b_{\d_1}$.

Define distribution $q_{k+1}$ over  $\Xc_b^{k+1}$ as the $(k+1)$-th order distribution of the quantized process $[X_1]_b, \ldots, [X_n]_b$. That is, for $a^{k+1}\in\Xc_b^{k+1}$,
\begin{align}
q_{k+1}(a_{k+1}|a^k)=\P([X_{k+1}]_b=a_{k+1}|[X^k]_b=a^k),\label{eq:q-k+1-source}
\end{align}
and
\begin{align}
q_{k}(a^{k})=\sum_{a_{k+1}\in\Xc_b}q_{k+1}(a^{k+1})=\P([X^k]_b=a^k).\label{eq:q-k-source}
\end{align}
Also define distributions $\hat{q}_{k+1}^{(1)}$ and $\hat{q}_{k+1}^{(2)}$ as the empirical distributions induced by $\Xh^n$ and $[X^n]_b$, respectively. In other words, $\hat{q}_{k}^{(1)}(a^{k})=\hat{p}^{(k)}(a^k|\Xh^n)$, $\hat{q}_{k}^{(2)}(a^{k})=\hat{p}^{(k)}(a^k|[X^n]_b),$ and
\begin{align}
\hat{q}_{k+1}^{(1)}(a_{k+1}|a^k)={\hat{q}_{k+1}^{(1)}(a^{k+1})\over \hat{q}_{k}^{(1)}(a^{k}) }={\hat{p}^{(k+1)}(a^{k+1}|\Xh^n)\over \hat{p}^{(k)}(a^k|\Xh^n)},
\end{align}
and
\begin{align}
\hat{q}^{(2)}_{k+1}(a_{k+1}|a^k)={\hat{q}_{k+1}^{(2)}(a^{k+1})\over \hat{q}_{k}^{(2)}(a^{k}) }={\hat{p}^{(k+1)}(a^{k+1}|[X^n]_b)\over \hat{p}^{(k)}(a^k|[X^n]_b)}.
\end{align}

As the first step we would like to prove that ${1\over b}  \hat{H}_{k}(\Xh^n)\leq  \bar{d}_k(\Xbbf)+\d$. Using the definitions above, we have
\begin{align}\label{eq:simplifyprobterm1}
\sum_{a^{k+1}\in\Xc_b^{k+1}} &w_{a^{k+1}} \hat{p}^{(k+1)}(a^{k+1}|\Xh^n)\nonumber\\
&= \sum_{a^{k+1}\in\Xc_b^{k+1}}\hat{p}^{(k+1)}(a^{k+1}|\Xh^n) \log{1\over q_{k+1}(a_{k+1}|a^k)}  \nonumber \\
&= \sum_{a^{k+1}\in\Xc_b^{k+1}} \hat{p}^{(k+1)}(a^{k+1}|\Xh^n)   \log{\hat{q}_{k+1}^{(1)}(a_{k+1}|a^k)\over q_{k+1}(a_{k+1}|a^k))}\nonumber\\
&\;\;\;\;\;\;\;+  \sum_{a^{k+1}\in\Xc_b^{k+1}}\hat{p}^{(k+1)}(a^{k+1}|\Xh^n)  \log{1\over \hat{q}_{k+1}^{(1)}(a_{k+1}|a^k)} \nonumber \\ 
&= \sum_{a^k}\hat{q}_{k}^{(1)}(a^k) D_{\rm KL}(\hat{q}_{k+1}^{(1)}(\cdot|a^k)\| q_{k+1}(\cdot|a^k)) +
 \hat{H}_{k}(\Xh^n).
\end{align}
Since $\Xh^n$ is the minimizer of \eqref{eq:Q-MAP}, we have
\begin{align}
\sum_{a^{k+1}\in\Xc_b^{k+1}} w_{a^{k+1}} \hat{p}^{(k+1)}(a^{k+1}|\Xh^n) \leq (\bar{d}_k(\Xbbf)+\d)b.
\end{align}
Combining this equation with \eqref{eq:simplifyprobterm1} and the fact that $D_{\rm KL}$ is always positive, we obtain
\begin{align}
{1\over b}  \hat{H}_{k}(\Xh^n)\leq  \bar{d}_k(\Xbbf)+\d.\label{eq:bd-H-hat-X-hat-Q-MAP}
\end{align}

As the second step of the proof we show that with high probability \begin{align}
{1\over b} \sum_{a^{k+1}\in\Xc_b^{k+1}} w_{a^{k+1}} \hat{p}^{(k+1)}(a^{k+1}|[X^n]_b) \leq \bar{d}_k(\Xbbf) +\d.
\end{align}
In other words, we would like to show that the vector $[X^n]_b=([X_1]_b, [X_2]_b, \ldots, [X_n]_b)$ satisfies the constraint of the optimization \eqref{eq:Q-MAP}. Following the same steps as those used in deriving \eqref{eq:simplifyprobterm1}, we get
\begin{align}
&\sum_{a^{k+1}\in\Xc_b^{k+1}} w_{a^{k+1}} \hat{p}^{(k+1)}(a^{k+1}|[X^n]_b)\nonumber\\
&= \sum_{a^k}\hat{q}_{k}^{(2)}(a^k) D_{\rm KL}(\hat{q}_{k+1}^{(2)}(\cdot|a^k)\| q_{k+1}(\cdot|a^k))+
 \hat{H}_{k}([X^n]_b).\label{eq:simplifyprobterm2}
\end{align}
Also, note that
\begin{align}
&\sum_{a^k}\hat{q}_{k}^{(2)}(a^k) D_{\rm KL}(\hat{q}_{k+1}^{(2)}(\cdot|a^k)\| q_{k+1}(\cdot|a^k))\nonumber\\
&=
\sum_{a^k}\hat{q}_{k}^{(2)}(a^k)\sum_{a_{k+1}} \hat{q}_{k+1}^{(2)}(a_{k+1}|a^k) \log {\hat{q}_{k+1}^{(2)}(a_{k+1}|a^k)\over  q_{k+1}(a_{k+1}|a^k)}\nonumber\\
&=\sum_{a^{k+1}}\hat{q}_{k+1}^{(2)}(a^{k+1})\Big( \log {\hat{q}_{k+1}^{(2)}(a^{k+1})\over  q_{k+1}(a^{k+1})}-\log {\hat{q}_{k}^{(2)}(a^{k})\over  q_{k}(a^{k})}\Big)\nonumber\\
&= D_{\rm KL}(\hat{q}_{k+1}^{(2)}\| q_{k+1})- D_{\rm KL}(\hat{q}_{k}^{(2)}\| q_{k}).\label{eq:D-KL-cond-regular}
\end{align}
Therefore, since $0\leq D_{\rm KL}(\hat{q}_{k+1}^{(2)}\| q_{k+1})- D_{\rm KL}(\hat{q}_{k}^{(2)}\| q_{k})\leq D_{\rm KL}(\hat{q}_{k+1}^{(2)}\| q_{k+1})$, from  \eqref{eq:simplifyprobterm2},
\begin{align}\label{eq:cost-vs-KL-distance}
\sum_{a^{k+1}\in\Xc_b^{k+1}} w_{a^{k+1}} \hat{p}^{(k+1)}(a^{k+1}|[X^n]_b)
\leq \hat{H}_{k}([X^n]_b) +D_{\rm KL}(\hat{q}_{k+1}^{(2)}\| q_{k+1}).
\end{align}
Given  $\d_2>0$,  define event $\Ec_1$ as
     \begin{equation}\label{eqdef:Ec1}
    \Ec_1\triangleq \{\|\hat{q}_{k+1}^{(2)}-q_{k+1}\|_1<\d_2\}.
    \end{equation}
Consider random vector $U^{k+1}$ distributed according to $\hat{q}_{k+1}^{(2)}$, which denotes the empirical distribution of $[X^n]_b$. Then, by definition, 
$\hat{H}_k([X^n]_b)= H(U_{k+1}|U^k)= H(U^{k+1})-H(U^k).$ 
Therefore,
\begin{align}
|\hat{H}_k([X^n]_b)-H([X_{k+1}]_b|[X^k]_b)|&=|H(U^{k+1})-H(U^k)-H([X^{k+1}]_b)+H([X^k]_b)|\nonumber\\
&\leq |H(U^{k+1})-H([X^{k+1}]_b)|+|H(U^k)-H([X^k]_b)|
\end{align}
Conditioned on $\Ec_1$, $\|\hat{q}_{k}^{(2)}-q_{k}\|_1\leq \|\hat{q}_{k+1}^{(2)}-q_{k+1}\|_1\leq \d_2$, and therefore, from Lemma \ref{lemma:bd-TV-dist}, 
\begin{align}
|\hat{H}_k([X^n]_b)-H([X_{k+1}]_b|[X^k]_b)|\leq -2\d_2\log \d_2 +2\d_2(k+1)\log |\Xc_b|,
\end{align}
or
\begin{align}
|{\hat{H}_k([X^n]_b)\over b}-{H([X_{k+1}]_b|[X^k]_b)\over b}|\leq -{2\d_2\over b}\log \d_2 +({2(k+1)\log |\Xc_b|\over b})\d_2.\label{eq:bd-dist-Hh-H-emp}
\end{align}
Moreover, conditioned on $\Ec_1$,  since  $\|\hat{q}_{k+1}^{(2)}-{q}_{k+1}\|_1\leq \d_2$ and $\hat{q}_{k+1}^{(2)} \ll {q}_{k+1}$, from Lemma \ref{lemma:dist-KL-vs-L1}, we have
\begin{align}
D(\hat{q}_{k+1}^{(2)}\|{q}_{k+1} )\leq -\d_2\log \d_2 +\d_2(k+1)\log |\Xc_b|-\d_2 \log q_{\min},
\end{align}
where   
\begin{align}
q_{\min}=\min_{u^{k+1}\in\Xc_b^{k+1}: q_{k+1}(u^{k+1})\neq 0} \P([X^{k+1}]_b=u^{k+1})\geq f_{k+1} |\Xc_b|^{-(k+1)}.
\end{align} Therefore,
\begin{align}
D(\hat{q}_{k+1}^{(2)}\|{q}_{k+1} )&\leq -\d_2\log \d_2 +\d_2(k+1)\log |\Xc_b|\nonumber\\
&\;\;\;-\d_2\log f_{k+1} +\d_2(k+1) \log |\Xc_b|,
\end{align}
or
\begin{align}
{D(\hat{q}_{k+1}^{(2)}\|{q}_{k+1} )\over b}&\leq -{\d_2\over b}(\log \d_2 +\log f_{k+1}) +({2(k+1)\log |\Xc_b|\over b})\d_2.\label{eq:bd-D-q2-lemma5-used}
\end{align}
 Hence, combining \eqref{eq:cost-vs-KL-distance}, \eqref{eq:bd-dist-Hh-H-emp} and \eqref{eq:bd-D-q2-lemma5-used}, it follows that, conditioned on $\Ec_1$, 
 \begin{align}
&{1\over b} \sum_{a^{k+1}\in\Xc_b^{k+1}} w_{a^{k+1}} \hat{p}^{(k+1)}(a^{k+1}|[X^n]_b) \nonumber\\
&\leq \bar{d}_k(\Xbbf)+\d_1 +({4(k+1)\log |\Xc_b|\over b})\d_2
 -{\d_2\over b}(3\log \d_2 +\log f_{k+1}).\label{eq:ub-Hhat-Xo-delat-1-delta-2}
 \end{align}
Choosing $\d_1=\d/2$ and $\d_2$  small enough such that
\begin{align}
&({4(k+1)\log |\Xc_b|\over b})\d_2
 -{\d_2\over b}(3\log \d_2 +\log f_{k+1})\leq {\d\over 2}. \label{eq:cond-on-delta2}
\end{align}
Note that while $|\Xc_b|$ grows exponentially in $b$, for all bounded sources, ${1\over b}\log |\Xc_b|<2$. Therefore, it is always possible to make sure that the above condition is satisfied for an appropriate choice of parameter $\d_2$. For this choice of parameters, from \eqref{eq:ub-Hhat-Xo-delat-1-delta-2}, conditioned on $\Ec_1$
\begin{equation}\label{eq:constraint}
{1\over b} \sum_{a^{k+1}\in\Xc_b^{k+1}} w_{a^{k+1}} \hat{p}^{(k+1)}(a^{k+1}|[X^n]_b) \leq \bar{d}_k(\Xbbf) +\d,
\end{equation}
and hence $[X^n]_b$ satisfies the constraint of the Q-MAP optimization described in \eqref{eq:Q-MAP}. Hence, since $\Xh^n$ is the minimizer of $\|Au^n-Y^m\|^2$, among all sequences that satisfy this constraint, we conclude that, conditioned on $\Ec_1$,
\begin{align}
\|A\Xh^n-Y^m\|&\leq \|A[X^n]_b-Y^m\|\nonumber\\
&= \|A([X^n]_b-X^n)\|\nonumber\\
&\leq \sigma_{\max}(A)\|X^n-[X^n]_b\|\nonumber\\
&\leq \sigma_{\max}(A)2^{-b}\sqrt{n}.\label{eq:error-Xh-vs-error-Xob}
\end{align}
Our goal is to use this equation to derive a bound for $\|\hat{X}^n- X^n\|$. The main challenge here is to find a lower bound for $\|A\Xh^n-Y^m\|$ in terms of $\|\hat{X}^n- X^n\|$. 
Given  $\d_3>0$ and $\tau>0$,    define set $\Cc_n$ and events $\Ec_2$ and $\Ec_3$  as
    \begin{equation}\label{eqdef:CCn}
     \Cc_n \triangleq \{u^n\in\Xc_b^n: {1\over nb}\ell_{\rm LZ}(u^n)\leq \bar{d}_k(\Xbbf)+2\d\},
    \end{equation}
     \begin{equation}\label{eqdef:E2n}
    \Ec_2 \triangleq \{\sigma_{\max}(A)<\sqrt{n}+2\sqrt{m}\},
    \end{equation}
and
     \begin{equation}\label{eqdef:E3n}
    \Ec_3 \triangleq \{\|A(u^n-X^n)\|\geq \|u^n-X^n\|\sqrt{(1-\tau)m}: \forall u^n\in\Cc_n\},
    \end{equation}
    respectively. We will prove the following:
    \begin{enumerate}
    \item $\Xh^n\in\Cc_n$, for $n$ large enough.
    \item $\P(\Ec_1\cap \Ec_2 \cap \Ec_3)$ converges to one as $n$ grows to infinity. 
    \end{enumerate}
    
For the moment we assume that  these two statements are true and complete the proof. Therefore, conditioned on $\Ec_1\cap\Ec_2\cap\Ec_3$, it follows from \eqref{eq:error-Xh-vs-error-Xob} that
 \begin{align}
\|\Xh^n-X^n\|\sqrt{(1-\tau)m} &\leq n(1+2\sqrt{m\over n})2^{-b}\nonumber\\
 &\leq 3n2^{-b},
\end{align}
where the last line follows form the fact that $m\leq n$. Therefore, conditioned on $\Ec_1\cap\Ec_2\cap\Ec_3$,
 \begin{align}
{1\over \sqrt{n}}\|\Xh^n-X^n\|\leq \sqrt{9 n\over (1-\tau)m2^{2b}}.\label{eq:error-Xh-vs-error-Xob-per-symb}
\end{align}

To prove that  $\Xh^n\in\Cc_n$, for $n$ large enough, note that, from \eqref{eq:bd-H-hat-X-hat-Q-MAP}, ${1\over b}\hat{H}_k(\Xh^n)\leq \bar{d}_k(\Xbbf)+\d$. On the other hand, from \eqref{eq:connections-LZ-Hk}, for our choice of parameter $b=b_n$,  for any  given $\d''>0$, for all $n$ large enough,
\begin{align}
{1\over n}\ell_{\rm LZ}(\Xh^n)\leq \hat{H}_k(\Xh^n)+\d''.
\end{align}
Therefore,  for all $n$ large enough,
\begin{align}
{1\over nb}\ell_{\rm LZ}(\Xh^n)\leq \bar{d}_k(\Xbbf)+\d+{\d''\over b}.
\end{align}
Choosing $\d''$ such that ${\d''\over b}\leq \d$ proves the desired result, \ie $\Xh^n\in\Cc_n$.

 Let $\tau=1-(\log n)^{-2r/(1+f)}$, where $f>0$ is a free parameter. For $b=b_n=\lceil r\log \log n\rceil$, $2^{2b}\geq (\log n)^{2r}$. Therefore, from \eqref{eq:error-Xh-vs-error-Xob-per-symb},  
 \begin{align}
{1\over \sqrt{n}}\|\Xh^n-X^n\|&\leq \sqrt{9 (\log n)^{2r\over 1+f} \over (1+\delta)\bar{d}_k(\Xbbf) (\log n)^{2r}} \nonumber\\
&= \sqrt{9 \over (1+\delta)\bar{d}_k(\Xbbf) (\log n)^{2rf \over 1+f}} .
\end{align}
Therefore, for any $\e>0$, $n$ large enough,  conditioned on $\Ec_1\cap\Ec_2\cap\Ec_3$,
\begin{align}
{1\over \sqrt{n}}\|\Xh^n-X^n\|\leq \e.
\end{align}

 To finish the proof we study the probability of $\Ec_1\cap\Ec_2\cap\Ec_3$. By Theorem \ref{thm:exp-rate-mixing}, there exists integer $g_{\delta_2}$,  only depending on the source distribution and $\d_2$ such that for $n>6(k+g_{\delta_2})/\d_2+k$,
    \begin{align}
    \P(\Ec_1^c)\leq 2^{c\d_2^2/8}(k+g_{\delta_2})n^{|\Xc_b|^k}2^{-{n\d_2^2 \over 8(k+g_{\delta_2})}},
\end{align}
where $c=1/(2\ln 2)$. Also,    as proved in \cite{CaTa05}, 
    \begin{align}
    \P(\Ec_2^c)\leq 2^{-m/2}.
    \end{align} Finally, from \eqref{eq:LZ-sequences}, the size of $\Cc_n$ can be upper-bounded as 
    \begin{align}
    |\Cc_n|\leq 2^{nb(\bar{d}_k(\Xbbf)+2\d)+1}.
    \end{align} 
 Now Lemma \ref{lemma:chi} combined with the union bound proves that,  for a fixed vector $X^n$,
    \begin{align}
    \P_A(\Ec_3^c)\leq  2^{nb(\bar{d}_k(\Xbbf)+2\d)+1} \ex^{{m\over 2}(\tau +\ln (1-\tau))},
    \end{align}
    where $\P_A$ reflects the fact that $[X^n]_b$ is fixed, and the randomness is in the generation of matrix $A$. 
For our choice of parameter $\tau$ combined with the Fubini's Theorem and the Borel Cantelli Lemma proves that $\P_{X^n}(\Ec_3^c)\to 0$, almost surely.


\subsection{Proof of Theorem \ref{thm:mainresult-Q-MAP-L}}\label{sec:proof-thm3}


The proof is very similar to the proof of Theorem \ref{thm:mainresult-Q-MAP} and follows the same logic. Similar to the proof of Theorem  \ref{thm:mainresult-Q-MAP}, for $\d_1>0$,  we assume that $n$ is larges enough such that 
\begin{align}
 {H([X_{k+1}]_b|[X^k]_b)\over b} \leq \bar{d}_k(\Xbbf) + \d_1.
\end{align}
Also,  given $\d_2>0$, $\d_3>0$ and $\tau>0$, we consider the events $\Cc_n$ and events $\Ec_2$ and $\Ec_3$ define in \eqref{eqdef:CCn}, \eqref{eqdef:E2n}, \eqref{eqdef:E3n}. Since $\Xh^n$ is a minimizer of \\ $\sum_{a^{k+1}\in\Xc_b^{k+1}} w_{a^{k+1}} \hat{p}^{(k)}(a^{k+1}|u^n) +{\lambda\over n^2}\|Au^n-Y^m\|^2$, we have
\begin{align}\label{eq:cost-Xhat-Xnb}
&\sum_{a^{k+1}\in\Xc_b^{k+1}} w_{a^{k+1}} \hat{p}^{(k+1)}(a^{k+1}|\Xh^n) +{\lambda\over n^2}\|A\Xh^n-Y^m\|^2\nonumber\\
&\leq \sum_{a^{k+1}\in\Xc_b^{k+1}} w_{a^{k+1}} \hat{p}^{(k+1)}(a^{k+1}|[X^n]_b) +{\lambda\over n^2}\|A[X^n]_b-Y^m\|^2\nonumber\\
&\leq \sum_{a^{k+1}\in\Xc_b^{k+1}} w_{a^{k+1}} \hat{p}^{(k+1)}(a^{k+1}|[X^n]_b) +{\lambda(\sigma_{\max}(A))^2\over n^2}\|[X^n]_b-X^n\|^2\nonumber\\
&\leq \sum_{a^{k+1}\in\Xc_b^{k+1}} w_{a^{k+1}} \hat{p}^{(k+1)}(a^{k+1}|[X^n]_b) +{\lambda(\sigma_{\max}(A))^22^{-2b}\over n}.
\end{align}
Define distribution $q_{k+1}$,  $\hat{q}_{k+1}^{(1)}$ and $\hat{q}_{k+1}^{(2)}$  over  $\Xc_b^{k+1}$ as in the proof of Theorem \ref{thm:mainresult-Q-MAP}. Then, given $\d>0$, from \eqref{eq:bd-dist-Hh-H-emp} and \eqref{eq:ub-Hhat-Xo-delat-1-delta-2},  we set $\d_1=\d/2$ and $\d_2$ small enough such that \eqref{eq:cond-on-delta2} is satisfied. Then following the same steps as the ones that led to \eqref{eq:constraint} we obtain that conditioned on $\Ec_1$ (defined in \eqref{eqdef:Ec1}), 
\begin{align}
{1\over b} \hat{H}_k([X^n]_b)\leq \bar{d}_k(\Xbbf)+\d.
\end{align}
Hence, we have
\begin{align}
\hat{H}_k(\Xh^n) +{\lambda\over n^2}\|A\Xh^n-Y^m\|^2 \leq b(\bar{d}_k(\Xbbf)+\d )  +{\lambda(\sigma_{\max}(A))^22^{-2b}\over n}.\label{eq:ub-cost-L-QMAP}
\end{align}
Since both terms on the left hand side of \eqref{eq:ub-cost-L-QMAP} are  positive, each of them should be smaller than the bound on the right hand side, \ie
\begin{align}\label{eq:upper-bd-Hhat-k}
{1\over b}\hat{H}_k(\Xh^n) &\;\leq\; \bar{d}_k(\Xbbf)+\d  +{\lambda(\sigma_{\max}(A))^22^{-2b}\over bn},
\end{align}
and 
\begin{align}\label{eq:upper-bd-meas-error}
{\lambda\over b n^2}\|A\Xh^n-Y^m\|^2&\;\leq\; \bar{d}_k(\Xbbf)+\d   +{\lambda(\sigma_{\max}(A))^22^{-2b}\over bn}.
\end{align}
Since $\l=\l_n=(\log n)^{2r}$ and $b=b_b=\lceil  r\log\log n\rceil$,  $\lambda 2^{-2b}\leq 1$, and hence, conditioned on $\Ec_2$,
\begin{align}
{\lambda(\sigma_{\max}(A))^22^{-2b}\over bn}\leq {(\sqrt{n}+2\sqrt{m})^2\over nb}\leq {9\over b}.
\end{align}
Therefore, since $b_n\to\infty$, as $n\to\infty$, for all $n$ large enough, conditioned on $\Ec_2$,
\begin{align}
{\lambda(\sigma_{\max}(A))^22^{-2b}\over bn}\leq \d,
\end{align}
and  therefore, from \eqref{eq:upper-bd-Hhat-k} and \eqref{eq:upper-bd-meas-error}, 
\begin{align}\label{eq:upper-bd-Hhat-k-simple}
{1\over b}\hat{H}_k(\Xh^n) &\;\leq\; \bar{d}_k(\Xbbf)+2\d,
\end{align}
and 
\begin{align}\label{eq:upper-bd-meas-error-simple}
{\lambda\over b n^2}\|A\Xh^n-Y^m\|^2&\;\leq\; \bar{d}_k(\Xbbf)+2\d.
\end{align}
Therefore, choosing $\d_3=3\d$, conditioned on $\Ec_1\cap\Ec_2\cap\Ec_3$,  $\Xh^n\in\Cc_n$. Finally, from \eqref{eq:upper-bd-meas-error} we have
    \begin{align}
     {\lambda(1-\tau)m\over n^2 b} \|\Xh^n-X^n\|^2 \leq  \bar{d}_k(\Xbbf)+2\d,
    \end{align}
    or 
      \begin{align}\label{eq:bound-error-triangle}
     {1\over\sqrt{ n}} \|\Xh^n-X^n\| &\leq  \sqrt{ (\bar{d}_k(\Xbbf)+2\d)bn\over \l (1-\tau)m}.
    \end{align}
    which proves that, for our set of parameters,  conditioned on $\Ec_1\cap\Ec_2\cap\Ec_3$, ${1\over \sqrt{n}}\|X^n-\Xh^n\|$ can be made arbitrary small. Setting of parameter $\tau$ and proving that $\P(\Ec_1\cap\Ec_2\cap\Ec_3)$ converges to one can be done exactly as it was done in the proof of Theorem \ref{thm:mainresult-Q-MAP}.


\subsection{Proof of Theorem \ref{thm:PGD-analysis-noisy}}\label{sec:proof-thm6}
As argued in the proof of Theorem \ref{thm:mainresult-Q-MAP}, given $\d>0$, there exists $b_{\d}$, such that for $b>b_{\d}$, 
\begin{align}
{H([X_{k+1}]_b|[X^k]_b)\over b}\leq \bar{d}_k(\Xbbf)+{\d\over 2}.\label{eq:conv-dk-cond-H}
\end{align}
In the rest of the proof we assume that $n$ is large enough so that $b=b_n> b_{\d}$. Define distributions $q_{k+1}$ and $\hat{q}_{k+1}$ over $\Xc_b^{k+1}$ as follows. Let  $q_{k+1}$ and $\hat{q}_{k+1}$ denote the distribution of $[X^{k+1}]_b$,  and the empirical distribution of $[X^n]_b$, respectively. From \eqref{eq:simplifyprobterm2} and \eqref{eq:D-KL-cond-regular}, it follows that 
\begin{align}
\sum_{a^{k+1}\in\Xc_b^{k+1}} w_{a^{k+1}} \hat{p}^{(k+1)}(a^{k+1}|[X^n]_b)\leq \hat{H}_k([X^n]_b)+D_{\rm KL}(\hat{q}_{k+1}\| q_{k+1})\label{eq:bd-dist-Hhat-W-Phat}
\end{align}

Define the  events $\Ec_1$ and $\Ec_2$ as 
  \begin{align}
    \Ec_1=\{\sigma_{\max}(A)<\sqrt{n}+2\sqrt{m}\},
    \end{align}
    and
     \begin{align}
    \Ec_2=\{\|\hat{q}_{k+1}-q_{k+1}\|_1<{\d'}\},
    \end{align}
     where $\d'>0$ is selected such that 
     \begin{align}
     {1\over b}(-2\d'\log {\d'} +2\d'(k+1)\log |\Xc_b|)\leq {\d\over 4},\label{eq:delta-p-vs-delta}
     \end{align}
     and 
     \begin{align}
		{1\over b}(-\d'\log {\d'}-\delta' \log f_{k+1} +2\d'(k+1)\log |\Xc_b|)\leq {\d\over 4},\label{eq:delta-p-vs-delta-2}
     \end{align}
     for all $b$ large enough. This is always possible, since $ (-2\d'\log {\d'})/b$ is a decreasing function of $b$ and  $\log|\Xc_b|/b$ can be upper-bounded by a constant not depending on $b$. Hence, by picking $\delta'$ small enough both \eqref{eq:delta-p-vs-delta} and \eqref{eq:delta-p-vs-delta-2} hold.

Given this choice of parameters, from \eqref{eq:bd-dist-Hh-H-emp}, conditioned on $\Ec_2$, we have
\begin{align}
{1\over b}|\hat{H}_k([X^n]_b)-H([X_{k+1}]_b|[X^k]_b)|\leq  {\d\over 4}.
\end{align}
Also, from Lemma \ref{lemma:dist-KL-vs-L1}, and \eqref{eq:bd-D-q2-lemma5-used}, conditioned on $\Ec_2$,
\begin{align}
{1\over b}D(\hat{q}_{k+1}\|q_{k+1} )&\leq -{\d'\over b}(\log \d'+\log f_{k+1}) +({2(k+1)\log |\Xc_b|\over b})\d'.
\end{align}
Therefore, from \eqref{eq:delta-p-vs-delta-2}, 
\begin{align}
{1\over b}D(\hat{q}_{k+1}\|q_{k+1} )&\leq {\d\over 4}.
\end{align}
Conditioned on $\Ec_2$, from  \eqref{eq:conv-dk-cond-H} and \eqref{eq:bd-dist-Hhat-W-Phat}, it follows that  
\begin{align}
{1\over b}\sum_{a^{k+1}\in\Xc_b^{k+1}} w_{a^{k+1}} \hat{p}^{(k+1)}(a^{k+1}|[X^n]_b)&\leq {\hat{H}_k([X^n]_b)\over b}+{1\over b}D_{\rm KL}(\hat{q}_{k+1}\| q_{k+1})\nonumber\\
&\leq {H([X_{k+1}]_b|[X^k]_b)\over b}+{\d\over 4}+{\d\over 4}\nonumber\\
&\leq  \bar{d}_k(\Xbbf)+{\d\over 2}+{\d\over 4}+{\d\over 4}=\bar{d}_k(\Xbbf)+\d,
\end{align}
which implies that $[X^n]_b\in\Fc_o$. Since $\Xh^n(t+1)$ is the solution of \eqref{eq:PGD-update}, automatically, $\Xh^n(t+1)\in\Fc_o$. Also, as we just proved, conditioned on $\Ec_2\cap\Ec_3$, $[X^n]_b\in\Fc_o$ as well. Therefore,  conditioned on $\Ec_2\cap\Ec_3$,
\begin{align}
\|\Xh^n(t+1)-S^n(t+1)\|\leq \|[X^n]_b-S^n(t+1)\|,
\end{align}
or equivalently
\begin{align}
\|\Xh^n(t+1)-[X^n]_b+[X^n]_b-S^n(t+1)\|&\leq \|[X^n]_b-S^n(t+1)\|.\label{eq:add-sub-Xon}
\end{align}
Raising both sides of \eqref{eq:add-sub-Xon} to power two and canceling out the common term from both sides, we derive
\begin{align}
\|\Xh^n(t+1)-[X^n]_b\|^2 +2\langle \Xh^n(t+1)-[X^n]_b,[X^n]_b-S^n(t+1) \rangle &\leq 0.
\end{align}
If we plug in the expression for $S^n(t+1)$ we obtain
\begin{align}
\|\Xh^n(t+1)-[X^n]_b\|^2 \leq& \;2\langle \Xh^n(t+1)-[X^n]_b,-[X^n]_b+S^n(t+1) \rangle\nonumber\\
&\hspace{-2cm}=\;2\langle \Xh^n(t+1)-[X^n]_b,-[X^n]_b+\Xh^n(t)+\mu A^T(Y^m-A\Xh^n(t)) \rangle\nonumber\\
&\hspace{-2cm}=\;2\langle \Xh^n(t+1)-[X^n]_b,\Xh^n(t)-[X^n]_b \rangle\nonumber\\
&\hspace{-2cm}\;\;\;\;-2\mu \langle \Xh^n(t+1)-[X^n]_b,A^TA(\Xh^n(t)-X^n) \rangle\nonumber\\
&\hspace{-2cm}\;\;\;\;+2\mu \langle \Xh^n(t+1)-[X^n]_b,A^TZ^m\rangle\nonumber\\
&\hspace{-2cm}=\;2\langle \Xh^n(t+1)-[X^n]_b,\Xh^n(t)-[X^n]_b \rangle\nonumber\\
&\hspace{-2cm}\;\;\;\;-2\mu \langle \Xh^n(t+1)-[X^n]_b,A^TA(\Xh^n(t)-[X]_b^n) \rangle\nonumber\\
&\hspace{-2cm}\;\;\;\;+2\mu \langle \Xh^n(t+1)-[X^n]_b,A^TA(X^n-[X]_b^n) \rangle\nonumber\\
&\hspace{-2cm}\;\;\;\;+2\mu \langle \Xh^n(t+1)-[X^n]_b,A^TZ^m\rangle\label{eq:Et+1-vs-Et-step1}
\end{align}
Define
\begin{align}
E^n(t) \triangleq \|\Xh^n(t+1)-[X^n]_b\|,
\end{align}
and 
\begin{align}
\Et^n(t) \triangleq {E^n(t)\over \|E^n(t)\|}.
\end{align}
Then,  it follows from \eqref{eq:Et+1-vs-Et-step1} that 
\begin{align}
\|E^n(t+1)\| \leq&\;2\langle \Et^n(t+1), \Et^n(t)\rangle \|E^n(t)\| -2\mu \langle \Et^n(t+1),A^TA \Et^n(t) \rangle \|E^n(t)\| \nonumber\\
&+2\mu \langle \Et^n(t+1),A^TA(X^n-[X]_b^n) \rangle\nonumber\\
&+2\mu \langle \Et^n(t+1),A^TZ^m\rangle\nonumber\\
=&\;2\Big(\langle \Et^n(t+1), \Et^n(t)\rangle -\mu \langle  A\Et^n(t+1),A \Et^n(t) \rangle \Big)\|E^n(t)\| \nonumber\\
&+2\mu \langle \Et^n(t+1),A^TA(X^n-[X]_b^n) \rangle\nonumber\\
&+2\mu \langle \Et^n(t+1),A^TZ^m\rangle \nonumber\\
\leq &\;2\Big(\langle \Et^n(t+1), \Et^n(t)\rangle -\mu \langle  A\Et^n(t+1),A \Et^n(t) \rangle \Big)\|E^n(t)\| \nonumber\\
&+2\mu\sigma_{\max}(A^TA)\|X^n-[X]_b^n\|\nonumber\\
&+2\mu \langle \Et^n(t+1),A^TZ^m\rangle. \label{eq:Et+1-vs-Et-step2}
\end{align}
Note that for a fixed $X^n$, $\Et^n(t)$ and $\Et^n(t+1)$ can only take a finite number of different values. Let $\Sc_e$ denote the set of all possible normalized error vectors. That is,
\begin{align}
\Sc_e \triangleq \left\{{u^n-v^n \over \|u^n-v^n\|}: u^n,v^n\in\Fc_o \right\}.
\end{align}
Clearly, $\Et^n(t)$ and $\Et^n(t+1)$ are both members of $\Sc_e$.
Define event $\Ec_3$ as follows
\begin{align}
\Ec_3\triangleq \Big\{ \langle u^n, v^n \rangle -{1\over m} \langle  Au^n,A v^n\rangle \leq 0.45 :\; \forall \; (u^n,v^n)\in\Sc_e^2\Big\}.
\end{align}
Conditioned on $\Ec_1\cap \Ec_2\cap \Ec_3$, it follows  from \eqref{eq:Et+1-vs-Et-step2} that 
\begin{align}
\|E^n(t+1)\| &\leq 0.9 \|E^n(t)\|+{2 (\sqrt{n}+2\sqrt{m})^2\over  m}(2^{-b}\sqrt{n})\nonumber\\
&\;\;\;+2\mu \langle \Et^n(t+1),A^TZ^m\rangle. \label{eq:Et+1-vs-Et-noisy-step3}
\end{align}
The only remaining term on the right hand side of  \eqref{eq:Et+1-vs-Et-noisy-step3}  is $2\mu \langle \Et^n(t+1),A^TZ^m\rangle$. To upper bound this term, we employ Lemma \ref{lemma:gaussian-vectors}. Let $A_i^m$, $i=1,\ldots,n$, denote the $i$-th column of matrix $A$. Then, 
\begin{align}
A^TZ^m=\left[\begin{array}{c}
\langle A_1^m,Z^m \rangle\\
\langle A_2^m,Z^m \rangle\\
\vdots\\
\langle A_n^m,Z^m \rangle\\
\end{array}
\right],
\end{align}
and 
\begin{align}
\langle \Et^n(t+1),A^TZ^m\rangle =\sum_{i=1}^n \Et_i(t+1)  \langle A_i^m,Z^m \rangle.
\end{align}
By Lemma \ref{lemma:gaussian-vectors}, for each $i$, $\langle A_i^m,Z^m \rangle$ is distributed as $\|Z^m\| G_i $, where $G_i$ is a standard normal distribution independent of $\|Z^m\|$. Therefore, since the columns of matrix $A$ are also independent, overall $\langle \Et^n(t+1),A^TZ^m\rangle$ is distributed as 
\begin{align}
\|Z^m\|\sum_{i=1}^n \Et_i(t+1) G_i,
\end{align}
where $G_i$ are i.i.d.~standard normal independent of $\|Z^m\|$.  
Given, $\tau_1>0$ and $\tau_2>0$, define events $\Ec_4$ and $\Ec_5$ as follows:
\begin{align}
\Ec_4\triangleq \{{1\over m}\|Z^m\|^2\leq (1+\tau_1)\sigma^2\},
\end{align}
and 
\begin{align}
\Ec_5\triangleq\{ |\langle\tilde{e}^n,G^n\rangle|^2\leq 1+\tau_2: \;\forall \tilde{e}^n\in\Sc_e\}.
\end{align}
By Lemma \ref{lemma:chi},
\begin{align}
\P(\Ec_4^c)&=\P\Big({1\over \sigma^2}\|Z^m\|^2> (1+\tau_1)m\Big)\nonumber\\
&\leq {\rm e} ^{-\frac{m}{2}(\tau_1 - \ln(1+ \tau_1))}.
\end{align}
For a fixed vector $\tilde{e}^n$, $\langle\tilde{e}^n,G^n\rangle =\sum_{i=1}^n \tilde{e}_iG_i$ has a standard normal distribution and therefore, by letting $m=1$ in Lemma \ref{lemma:chi}, it follows that
\begin{align}
\P(|\langle\tilde{e}^n,G^n\rangle|^2 >1+\tau_2 )\leq {\rm e} ^{-0.5(\tau_2 - \ln(1+ \tau_2))}.
\end{align}
Hence, by the union bound, 
\begin{align}
\P(\Ec_4^c)&\leq |\Sc_e|{\rm e} ^{-0.5(\tau_2 - \ln(1+ \tau_2))}\nonumber\\
&\leq  2^{nb(\bar{d}_k(\Xbbf)+2\d)}{\rm e} ^{-0.5(\tau_2 - \ln(1+ \tau_2))},
\end{align}
where the last inequality follows from \eqref{eq:size-Se-vs-Fo} and \eqref{eq:size-Fo}.
But, for $\tau_2>7$, 
\begin{align}
{\rm e} ^{-0.5(\tau_2 - \ln(1+ \tau_2))}\leq 2^{-0.5 \tau_2},
\end{align}
which implies that for $\tau_2>7$,
\begin{align}
\P(\Ec_5^c)\leq  2^{nb(\bar{d}_k(\Xbbf)+2\d)-0.5\tau_2}.
\end{align}
Setting  
\begin{align}
\tau_2=2nb(\bar{d}_k(\Xbbf)+3\d)-1
\end{align}
 ensures that
\begin{align}
\P(\Ec_5^c)\leq   2^{-\d bn+0.5},
\end{align}
which converges to zero as $n$ grows to infinity. Finally, setting $\tau_1=1$, conditioned on $\Ec_4\cap\Ec_5$, we have
\begin{align}
2\mu \langle \Et^n(t+1),A^TZ^m\rangle& ={2\over m} \langle \Et^n(t+1),A^TZ^m\rangle \nonumber\\
&\leq {1\over m}\sqrt{m(1+\tau_1)\sigma^2(1+\tau_2)}\nonumber\\
&= \sqrt{(2m)\sigma^2 (2nb(\bar{d}_k(\Xbbf)+3\d))\over m^2}\nonumber\\
&= { \sigma\over 2}\sqrt{nb(\bar{d}_k(\Xbbf)+3\d)\over m}.\label{eq:bd-noisy-proj}
\end{align}
Combining  
 \eqref{eq:bd-noisy-proj} and \eqref{eq:Et+1-vs-Et-noisy-step3} yields the desired upper bound on the last term in \eqref{eq:Et+1-vs-Et-noisy-step3}.


To finish the proof we need to show that $\P(\Ec_1\cap\Ec_2\cap\Ec_3)$ also approaches one, as $n$ grows without bound.  Reference \cite{CaTa05} proves that
\begin{align}
\P(\Ec_1^c)\leq 2^{-m/2}.
\end{align}

By  Theorem \ref{thm:exp-rate-mixing}, there  exists integer $g_{\d'}$, depending only on $\d'$ and the source distribution, such that for any $n>6(k+g_{\d'})/(b\d)+k$,
\begin{align}
\P(\Ec_3^c)\leq 2^{c\d'^2/8} (k+g_{\d'}+1)n^{|\Xc_b|^{k+1}}2^{-{nc \d'^2 \over 8(k+g_{\d'}+1)}},
\end{align}
where $c=1/(2\ln 2)$. This proves that for our choice of parameters, $\P(\Ec_2^c)$ converges to zero. 

In the rest of the proof we bound $\P(\Ec_3^c)$. From Corollary \ref{cor:bound-45}, for any $u^n,v^n\in\Sc_e$,
\begin{align}
\P\Big({1\over m}\langle Au^n,Av^n\rangle-\langle u^n,v^n\rangle\leq -0.45\Big)\leq 2^{-0.05m}.
\end{align}
Therefore, by the union bound, 
\begin{align}
\P(\Ec_3^c)&=\P\Big({1\over m}\langle Au^n,A\tilde{e}^n(t)\rangle-\langle u^n,\tilde{e}^n(t)\rangle\leq -0.45: {\rm for}\;{\rm some}\;(u^n,v^n)\in\Sc_e^2\Big)\nonumber\\
&\leq |\Sc_e|^2 2^{-0.05m}.\label{eq:ub-Ec1-union}
\end{align}
Note that 
\begin{align}
|\Sc_e|\leq |\Fc_o|^2.\label{eq:size-Se-vs-Fo}
\end{align}
In the following we derive an upper bound on the size of $\Fc_o$.
For any $u^n\in\Fc_o$, by definition, we have
\begin{align}
\sum_{a^{k+1}\in\Xc_b^{k+1}} w_{a^{k+1}} \hat{p}^{(k+1)}(a^{k+1}|u^n)\leq b(\bar{d}_k(\Xbbf)+\d).\label{eq:compare-cost-linear-W}
\end{align}
Let $\hat{q}_{k+1}^{(u)}=\hat{p}^{(k+1)}(\cdot|u^n)$ denote the $(k+1)$-th order empirical distributions of $u^n$. Following the steps used in deriving \eqref{eq:simplifyprobterm1}, it follows that 
\begin{align}\label{eq:bd-HXt}
&\sum_{a^{k+1}\in\Xc_b^{k+1}} w_{a^{k+1}} \hat{p}^{(k+1)}(a^{k+1}|u^n)\nonumber\\
&=
 \sum_{a^k}\hat{q}_{k}^{(u)}(a^k) D_{\rm KL}(\hat{q}_{k+1}^{(u)}(\cdot|a^k)\| q_{k+1}(\cdot|a^k))+
 \hat{H}_{k}(u^n).
\end{align}
Since all the terms on the right hand side or \eqref{eq:bd-HXt} are positive, it follows from \eqref{eq:compare-cost-linear-W} that
\begin{align}
 \hat{H}_{k}(u^n)&\leq  b(\bar{d}_k(\Xbbf)+\d). \label{eq:bd-Hhat-Xh-Xo}
\end{align}
On the other hand, given our choice of quantization level $b$, for $n$ large enough, for any $v^n\in\Xc_b^n$,
\begin{align}
{1\over nb}\ell_{\rm LZ}(v^n)&\leq  {1\over b} \hat{H}_k(v^n)+\d.\label{eq:UB-LZ}
\end{align}
Therefore,  for any $u^n\in\Fc_o$, from  \eqref{eq:bd-Hhat-Xh-Xo} and \eqref{eq:UB-LZ}, it follows that
\begin{align}
{1\over nb}\ell_{\rm LZ}(u^n)&\leq  {1\over b} \hat{H}_k(u^n)+\d \nonumber\\
&\leq \bar{d}_k(\Xbbf)+2\d.
\end{align}
Note that, from \eqref{eq:LZ-sequences}, we  have 
\begin{align}
|\Fc_o|\leq |\{v^n:\; {1\over nb}\ell_{\rm LZ}(v^n)\leq\bar{d}_k(\Xbbf)+2\d\}|\leq  2^{nb(\bar{d}_k(\Xbbf)+2\d)}.\label{eq:size-Fo}
\end{align}
Hence, from \eqref{eq:ub-Ec1-union}, 
\begin{align}
&\P(\Ec_3^c)\leq |\Fc_o|^4 2^{-0.05m}\leq 2^{ nb(\bar{d}_k(\Xbbf)+2\d)-0.05m}\leq 2^{-2nb \d}.\label{eq:bound-et-n}
\end{align}

\section{Conclusion}\label{sec:conclusions}
We studied the problem of estimating $X^n \in \mathbb{R}^n$ from $m$ response variables $Y^m = AX^n+Z^m$, under the assumption that the distribution of $X^n$ is known. We proposed a new approach optimization called Q-MAP for estimating $X^n$. The new  optimization satisfies   the following properties: (i) It applies to  generic classes of distributions, as long as they satisfy certain mixing conditions. (ii) Unlike  other Bayesian approaches, in the high-dimensional settings, the performance of the Q-MAP optimization can be characterized for generic distributions. Our analyses show  that, for certain distributions such as spike-and-slab, asymptotically, Q-MAP achieves the minimum normalized number of measurements (Whether Q-MAP achieves the optimal bound for other distributions  is still an open question.)
(iii) Projected gradient descent can be applied to approximate the solution of the optimization Q-MAP.  While the optimization involved in Q-MAP is non-convex, we  have characterized the performance of the corresponding PGD algorithm, under both noiseless and noisy settings.  Our analysis revealed that with slightly more measurements than Q-MAP, the PGD-based method  recovers $X^n$ accurately in the noiseless setting. 

\section{Funding}
This work was supported by the National Science Foundation  [CCF-1420575 to S.J., CCF-1420328 to A.M.].

\bibliographystyle{unsrt}
\bibliography{../myrefs}

\end{document}